\documentclass[apj, tighten, numberedappendix, twocolappendix, appendixfloats]{emulateapj}
\pdfoutput = 1

\usepackage{graphicx}
\usepackage{amsfonts,amsmath,amssymb}
\usepackage{url}
\usepackage{hyperref}
\usepackage[english]{babel}
\usepackage{natbib}

\bgroup\catcode`\#=12\egroup \newcommand{\revise}[1]{#1}
\bgroup\catcode`\#=12\egroup \newcommand{\newrevise}[1]{#1}

\begin{document}

\title{KIC 9246715: The Double Red Giant Eclipsing Binary With Odd Oscillations
\shorttitle{KIC 9246715}
\shortauthors{RAWLS, M. L., GAULME, P., MCKEEVER, J. ET AL.}}

\slugcomment{Accepted for publication in ApJ, 2015 December 31}

\author{
Meredith L. Rawls,\altaffilmark{1} Patrick Gaulme,\altaffilmark{2}$^{,}$\altaffilmark{1} Jean McKeever,\altaffilmark{1} Jason Jackiewicz,\altaffilmark{1} Jerome A. Orosz,\altaffilmark{3} \\ Enrico Corsaro,\altaffilmark{4}$^{,}$\altaffilmark{5}$^{,}$\altaffilmark{6} Paul Beck,\altaffilmark{4} Beno\^it Mosser,\altaffilmark{7} David W. Latham,\altaffilmark{8} and Christian A. Latham\altaffilmark{8}
}
\affil{\altaffilmark{1}Department of Astronomy, New Mexico State University, P.O. Box 30001, MSC 4500, Las Cruces, NM 88003, USA; mrawls@nmsu.edu \\
\altaffilmark{2}Apache Point Observatory, 2001 Apache Point Road, P.O. Box 59, Sunspot, NM 88349, USA \\
\altaffilmark{3}Department of Astronomy, San Diego State University, 5500 Campanile Drive, San Diego, CA 91945, USA \\
\altaffilmark{4}Laboratoire AIM, CEA/DSM {\textendash} CNRS {\textendash} Univ. Paris Diderot {\textendash} IRFU/SAp, Centre de Saclay, 91191 Gif-sur-Yvette Cedex, France \\
\altaffilmark{5}Instituto de Astrof\'isica de Canarias, 38205 La Laguna, Tenerife, Spain \\
\altaffilmark{6}Universidad de La Laguna, Departamento de Astrof\'isica, 38206 La Laguna, Tenerife, Spain \\
\altaffilmark{7}LESIA, Observatoire de Paris, PSL Research University, CNRS, Universit\'e Pierre et Marie Curie, Universit\'e Paris Diderot, 92195 Meudon, France \\
\altaffilmark{8}Harvard-Smithsonian Center for Astrophysics, 60 Garden Street, Cambridge, MA 02138, USA}

\begin{abstract}
We combine \emph{Kepler} photometry with ground-based spectra to present a comprehensive dynamical model of the double red giant eclipsing binary KIC 9246715. While the two stars are very similar in mass ($M_1 = 2.171\substack{+0.006 \\ -0.008} \ M_{\odot}$, $M_2 = 2.149\substack{+0.006 \\ -0.008} \ M_{\odot}$) and radius ($R_1 = 8.37\substack{+0.03 \\ -0.07} \ R_{\odot}$, $R_2 = 8.30\substack{+0.04 \\ -0.03} \ R_{\odot}$), an asteroseismic analysis finds one main set of solar-like oscillations with unusually \newrevise{low-amplitude, wide modes}. A second set of oscillations from the other star may exist, but this marginal detection is extremely faint. Because the two stars are nearly twins, KIC 9246715 is a difficult target for a precise test of the asteroseismic scaling relations, which yield $M = 2.17\pm0.14 \ M_{\odot}$ and $R = 8.26\pm0.18 \ R_{\odot}$. \newrevise{Both stars are consistent with the inferred asteroseismic properties, but we suspect the main oscillator is Star 2 because it is less active than Star 1.} We find evidence for stellar activity and modest tidal forces acting over the 171-day eccentric orbit, which are likely responsible for the essential lack of solar-like oscillations in one star and weak oscillations in the other. Mixed modes indicate the main oscillating star is on the secondary red clump (a core-He-burning star), and stellar evolution modeling supports this with a coeval history for a pair of red clump stars. This system is a useful case study and paves the way for a detailed analysis of more red giants in eclipsing binaries, an important benchmark for asteroseismology.
\end{abstract}

\bibliographystyle{apj}


\section{Introduction}\label{intro}


Mass and radius are often-elusive stellar properties that are critical to understanding a star's past, present, and future. Eclipsing binaries are the only astrophysical laboratories that allow for a direct measurement of these and other fundamental physical parameters. Recently, however, observing solar-like oscillations in stars with convective envelopes has opened a window to stellar interiors and provided a new way to measure global stellar properties. A pair of asteroseismic scaling relations use the Sun as a \revise{benchmark between} these oscillations and a star's effective temperature to yield mass and radius \citep{kje95,hub10,mos13}.

While both the mass and radius scaling relations are useful, it is important to test their validity. Recent work has investigated the radius relation by comparing the asteroseismic large-frequency separation $\Delta \nu$ and stellar radius between models and simulated data \citep[e.g.][]{ste09,whi11,mig13}, and by comparing asteroseismic radii with independent radius measurements such as interferometry or binary star modeling \citep[e.g.][]{hub11,hub12,sil12}. All of these find that radius estimates from asteroseismology are precise within a few percent, with greater scatter for red giants than main sequence stars. The mass scaling relation remains relatively untested. Most studies test the $\Delta\nu$ scaling with average stellar density and not the scaling of $\nu_{\rm{max}}$ (the asteroseismic frequency of maximum oscillation power) with stellar surface gravity, because the latter has a less-secure theoretical basis \revise{\citep{bel11}.} It is not yet possible to reliably predict oscillation mode amplitudes as a function of frequency \citep{chr12}. One study by \citet{fra13} did test both scaling laws with the red giant eclipsing binary KIC 8410637. They found good agreement between Keplerian and asteroseismic mass and radius, but a more recent analysis from \citet{hub14} indicates that the asteroseismic density of KIC 8410637 is underestimated by $\sim$7\,\% (1.8~$\sigma$, accounting for the density uncertainties), which results in an overestimate of the radius by $\sim$9\,\% (2.7~$\sigma$) and mass by $\sim$17\,\% (1.9~$\sigma$). Additional benchmarks for the asteroseismic scaling relations are clearly needed.

Evolved red giants are straightforward to characterize through pressure-mode solar-like oscillations in their convective zones, and red giant asteroseismology is quickly becoming an important tool to study stellar populations throughout the Milky Way \citep[for a review of this topic, see][]{cha13}. Compared to main-sequence stars, red giants oscillate with larger amplitudes and longer periods---several hours to days instead of minutes. Oscillations appear as spikes in the amplitude spectrum of a light curve that is sampled both frequently enough and for a sufficiently long duration. Therefore, observations from the \emph{Kepler} space telescope taken every 29.4 minutes (long-cadence) over many 90-day quarters are ideal for asteroseismic studies of red giant stars.

\emph{Kepler}'s primary science goal is to find Earth-like exoplanets orbiting sun-like stars \citep{bor10}. However, in addition to successes in planet-hunting and suitability for red giant asteroseismology, \emph{Kepler} is also incredibly useful for studies of eclipsing binary stars. \emph{Kepler} has discovered numerous long-period eclipsing systems from consistent target monitoring over several years \citep{prs11,sla11}. Eclipsing binaries are important tools for understanding fundamental stellar properties, testing stellar evolutionary models, and determining distances. When radial velocity curves exist for both stars in an eclipsing binary, along with a well-sampled light curve, the inclination is precisely constrained and a full orbital solution with masses and radii can be found. Kepler's third law applied in this way is the \emph{only} direct method for measuring stellar masses.

Taken together, red giants in eclipsing binaries (hereafter RG/EBs) that exhibit solar-like oscillations are ideal testbeds for asteroseismology. There are presently 18 known RG/EBs that show solar-like oscillations \citep{hek10,gau13,gau14,bec14,bec15} with orbital periods ranging from 19 to 1058 days, all in the \emph{Kepler} field of view.

In this paper, we present physical parameters for the unique RG/EB KIC 9246715 with a combination of dynamical modeling, stellar atmosphere modeling, and asteroseismology. KIC 9246715 contains two nearly-identical red giants in a 171-day eccentric orbit with a single main set of solar-like oscillations. A second set of oscillations, \newrevise{potentially} attributable to the other star, is marginally detected. In \S \ref{data}, we describe how we acquire and process photometric and spectroscopic data, and \S \ref{rvs} explains our radial velocity extraction process. In \S \ref{atm}, we disentangle each star's contribution to the spectra to perform stellar atmosphere modeling. We then present our final orbital solution and physical parameters for KIC 9246715 in \S \ref{model}. Finally, \S \ref{discuss} compares our results with global asteroseismology and discusses the connection among solar-like oscillations, stellar evolution, and effects such as star spots and tidal forces, as well as implications for future RG/EB studies.

\section{Observations}\label{data}

\subsection{\emph{Kepler} light curves}\label{kepler}
Our light curves are from the \emph{Kepler} Space Telescope in long-cadence mode (one data point every 29.4 minutes), and span 17 quarters---roughly four years---with only occasional gaps. These light curves are well-suited for red giant asteroseismology, as main sequence stars with convective envelopes oscillate too rapidly to be measured with \emph{Kepler} long-cadence data.

When studying long-period eclipsing binaries, it is important to remove instrumental effects in the light curve while preserving the astrophysically interesting signal. In this work, we prioritize preserving eclipses. Our detrending algorithm uses the simple aperture photometry (SAP) long-cadence \emph{Kepler} data for quarters 0--17. First, any observations with NaNs are removed, and observations from different quarters are put onto the same median level so that the eclipses line up. The out-of-eclipse portions of the light curve are flattened, which removes any out-of-eclipse variability. For eclipse modeling, we use only the portions of the light curve that lie within one eclipse duration of the start and end of each eclipse. This differs from the light curve processing needed for asteroseismology, \revise{which typically} ``fills'' the eclipses to minimize their effect on the power spectrum \citep{gau14}.

\begin{figure*}[h!]
\begin{center}
\includegraphics[width=1.8\columnwidth]{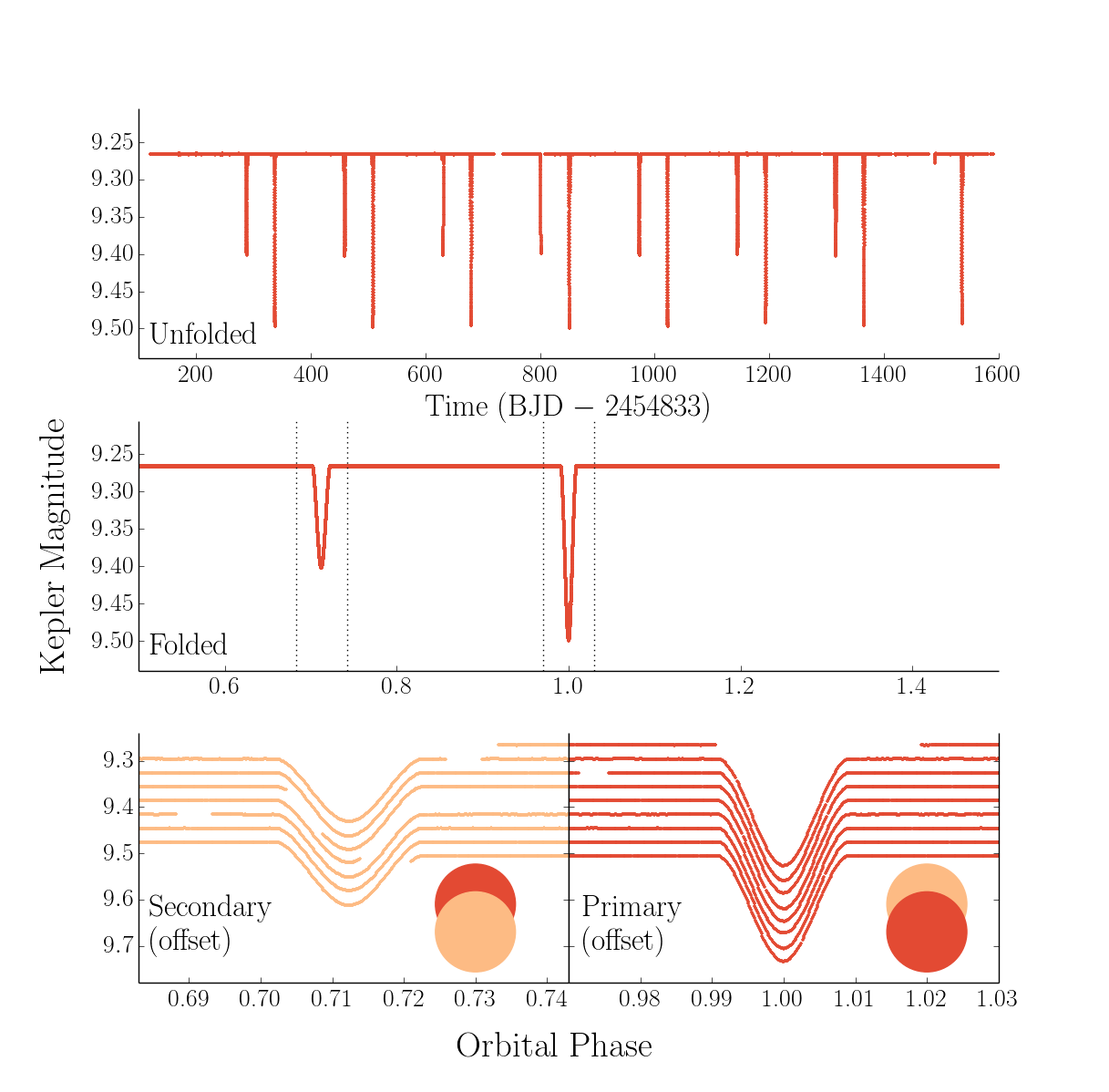}
\caption{\label{fig:keplerfig} \emph{Kepler} light curve of the eclipsing binary KIC 9246715 with out-of-eclipse points flattened. \emph{Top:} Detrended SAP flux over 17 quarters. The detrending process is described in Section \ref{kepler}. \emph{Middle:} Folded version of the above over one orbit. The dotted lines indicate the portion of the light curve used in subsequent modeling. \emph{Bottom:} A zoomed view of secondary and primary eclipses corresponding to the dotted lines above. To avoid overlaps, each observed eclipse is offset in magnitude from the previous one. The colored disks illustrate the eclipse configuration, with the red disk representing Star 1 and the yellow disk representing Star 2.
}
\end{center}
\end{figure*}

\begin{figure*}[h!]
\begin{center}
\includegraphics[width=1.7\columnwidth]{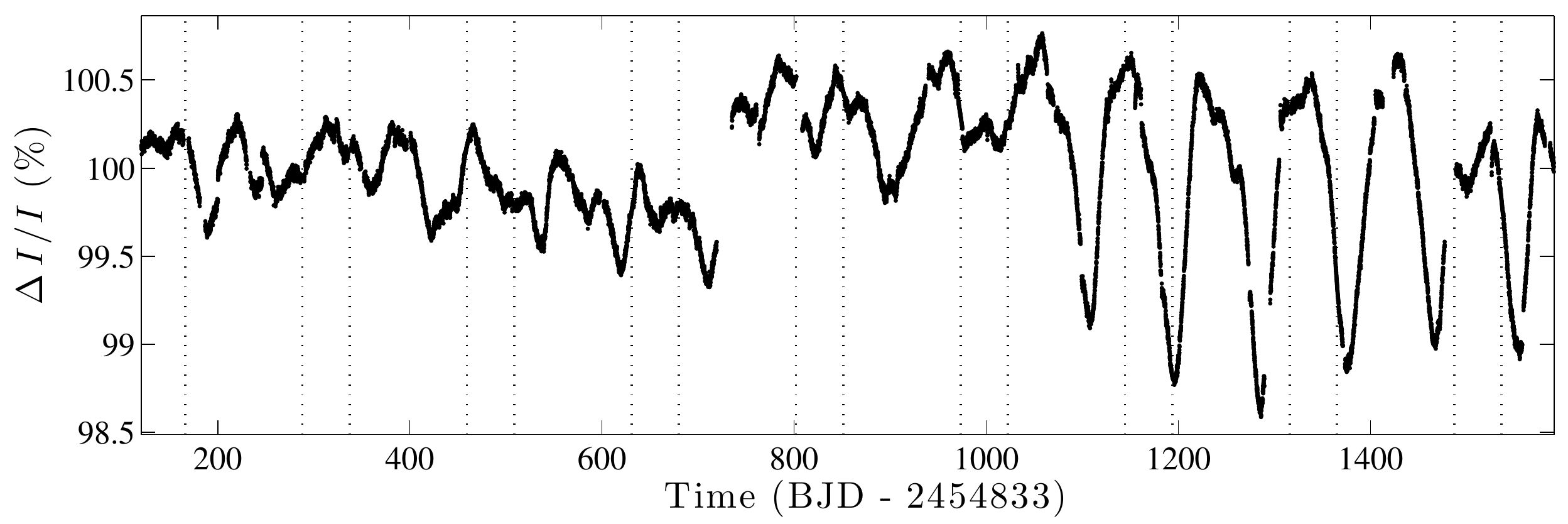}
\caption{\label{fig:lcfig2} \emph{Kepler} light curve of the eclipsing binary KIC 9246715 with eclipses removed, but retaining out-of-eclipse variability. The times of eclipses are indicated with dotted lines.
}
\end{center}
\end{figure*}

The processed light curve is presented in Figure \ref{fig:keplerfig}. The top panel shows the entire detrended light curve, while the middle and bottom panels indicate the regions near each eclipse used in this work. We adopt the convention that the ``primary'' eclipse is the deeper of the two, when Star 1 is eclipsing Star 2. The geometry of the system creates partial eclipses with different depths due to similarly-sized stars in an eccentric orbit viewed with an inclination less than 90 degrees. \revise{For comparison, we present the detrended light curve with eclipses removed in Figure \ref{fig:lcfig2}. The system shows out-of-eclipse photometric modulations on the order of $2 \%$.}

\subsection{Ground-based spectroscopy}\label{spectra}
We have a total of 25 high-resolution spectra from three spectrographs. At many orbital phases, prominent absorption lines show a clear double-lined signature when inspected by eye. We find that KIC 9246715 is an excellent target for obtaining radial velocity curves for both stars in the binary as the stellar flux ratio is close to unity. A long time span of observations was necessary due to the 171.277-day orbital period and visibility of the \emph{Kepler} field from the observing sites.

\subsubsection{TRES echelle from FLWO}\label{tres}
We obtained 13 high-resolution optical spectra from the Fred Lawrence Whipple Observatory (FLWO) 1.5-m telescope in Arizona using the Tillinghast Reflector Echelle Spectrograph (TRES) from 2012 March through 2013 April. The wavelength range for TRES is 3900--9100 \AA, and the resolution for the medium fiber used is 44,000. The spectra were extracted and blaze-corrected with the pipeline developed by \citet{buc10}.

\subsubsection{ARCES echelle from APO}\label{arces}
We also obtained ten high-resolution optical spectra from the Apache Point Observatory (APO) 3.5-m telescope in New Mexico using the Astrophysical Research Consortium Echelle Spectrograph (ARCES) from 2012 June through 2013 September. The wavelength range for ARCES is 3200--10,000 \AA \ with no gaps, and the average resolution is 31,000. We reduced the data using standard echelle reduction techniques and Karen Kinemuchi's ARCES cookbook (private communication)\footnote{\url{http://astronomy.nmsu.edu:8000/apo-wiki/wiki/ARCES} - ARCES Data Reduction Cookbook}.

\subsubsection{APOGEE spectra from APO}\label{apogee}
We finally obtained two near-IR spectra of KIC 9246715 from the Sloan Digital Sky Survey-III (SDSS-III) Apache Point Observatory Galactic Evolution Experiment (APOGEE) survey \citep{ala15}. The wavelength range for APOGEE is 1.5--1.7 $\mu$m with a nominal resolution of 22,500. The pair of spectra were reduced with the standard APOGEE pipeline, but not combined.

\subsubsection{Global wavelength solution}\label{wavelength}
Because the observations come from three different spectrographs at two different observatory sites, it is critical to apply a consistent wavelength solution that yields the same radial velocity zeropoint for all observations. This zeropoint is a function of the atmospheric conditions at the observatory and the instrument being used. Typically such a correction can be done with RV standard stars after a wavelength solution has been applied based on ThAr lamp observations. However, we lacked RV standard star observations, and some of the earlier ARCES observations had insufficiently frequent ThAr calibration images to arrive at a reliable wavelength solution. (We subsequently took ThAr images more frequently to address the latter issue.) To arrive at a consistent velocity zeropoint for all spectra, we use TelFit \citep{gul14} to generate a telluric line model of the O2 A-band (7595--7638 \AA) with $R = 31,000$ at STP. We then shift the ARCES and TRES spectra in velocity space using the broadening function technique (see Section \ref{bf}) so they all line up with the TelFit model. The shifts range from $-0.88$ to $2.18$ km s$^{-1}$, with the majority having a magnitude $< 0.3$ km s$^{-1}$.

\section{Radial velocities}\label{rvs}

\subsection{The broadening function}\label{bf}
To extract radial velocities from the spectra, we use the broadening function (BF) technique as outlined by \citet{ruc02}. In the simplest terms, the BF is a function that transforms any sharp-line spectrum into a Doppler-broadened spectrum. The BF technique involves solving a convolution equation for the Doppler broadening kernel $B$, $P(x) = \int B(x') T(x - x') dx'$, where $P$ is an observed spectrum of a binary and $T$ is a spectral template spanning the same wavelength window \citep{ruc15}. In practice, the BF can be used to characterize any deviation of an observed spectrum from an idealized sharp-line spectrum: various forms of line broadening, shifted lines due to Doppler radial velocity shifts, two sets of lines in the case of a spectroscopic binary, etc. The BF deconvolution is solved with singular value decomposition. This technique is generally preferred over the more familiar cross-correlation function (CCF), because the BF is a true linear deconvolution while the CCF is a non-linear proxy and is less suitable for double-lined spectra. The BF technique normalizes the result so that the velocity integral $\int B(v) dv = 1$ for an exact spectral match of the observed and template spectra. For this analysis, we adapt the IDL routines provided by Rucinski\footnote{\url{http://www.astro.utoronto.ca/\~rucinski/SVDcookbook.html}} into python\footnote{\url{https://github.com/mrawls/BF-rvplotter}}.

\begin{deluxetable}{cccccc}
\tablecolumns{6}
\tablewidth{0pt}
\tabletypesize{\footnotesize}
\tablecaption{Radial velocities for KIC 9246715 extracted from spectra with the broadening function technique.}
\centering
\tablehead{
\colhead{UTC} \vspace{-0.15cm} & \colhead{} & \colhead{} & \colhead{$v_1$} & \colhead{$v_2$} & \colhead{} \\
\colhead{} \vspace{-0.15cm} & \colhead{Midpoint$^{\rm{a}}$} & \colhead{Phase} & \colhead{} & \colhead{} & \colhead{Inst$^{\rm{b}}$} \\
\colhead{Date} & \colhead{} & \colhead{} & \colhead{($\rm{km \ s}^{-1}$)} & \colhead{($\rm{km \ s}^{-1}$)} & \colhead{} 
}
\startdata
2012-03-01 & 5988.047280 & 0.773 & $20.72(14)$  & $-29.88(14)$ & T \\
2012-03-11 & 5998.009344 & 0.831 & $34.91(14)$  & $-44.26(14)$ & T \\
2012-04-02 & 6020.026793 & 0.960 & $20.25(15)$  & $-29.77(15)$ & T \\
2012-05-08 & 6055.977358 & 0.170 & $-22.49(14)$  & $ 13.69(14)$ & T \\
2012-05-26 & 6073.937068 & 0.275 & $-26.35(14)$  & $ 17.53(14)$ & T \\
2012-06-02 & 6080.976302 & 0.316 & $-26.37(14)$  & $ 17.64(14)$ & T \\
2012-06-12 & 6090.904683 & 0.374 & $-25.55(15)$  & $ 16.67(15)$ & A \\
2012-06-27 & 6105.752943 & 0.460 & $-22.83(15)$  & $ 12.51(15)$ & A \\
2012-06-30 & 6108.894850 & 0.479 & $-21.01(14)$  & $ 12.20(14)$ & T \\
2012-07-24 & 6132.758456 & 0.618 & $-8.72(31)$  & $-0.55(32)$ & T \\
2012-08-26 & 6165.786902 & 0.811 & $29.96(15)$  & $-39.77(15)$ & A \\
2012-08-26 & 6165.947831 & 0.812 & $28.86(15)$  & $-41.26(15)$ & A \\
2012-08-27 & 6166.889910 & 0.817 & $33.01(15)$  & $-39.84(15)$ & A \\
2012-09-04 & 6174.917425 & 0.864 & $40.45(15)$  & $-48.07(15)$ & A \\
2012-09-05 & 6175.777945 & 0.869 & $39.85(14)$  & $-49.22(14)$ & T \\
2012-09-30 & 6200.689766 & 0.015 & $2.35(18)$  & $-11.53(20)$ & T \\ 
2012-10-24 & 6224.736100 & 0.155 & $-21.22(14)$  & $ 12.80(14)$ & T \\
2012-11-21 & 6252.572982 & 0.318 & $-26.39(14)$  & $ 17.67(14)$ & T \\
2013-04-02 & 6384.991673 & 0.091 & $-14.11(15)$  & $ 5.13(14)$ & T \\
2013-04-20 & 6402.975545 & 0.196 & $-23.98(15)$  & $ 15.28(15)$ & A \\
2013-06-13 & 6456.959033 & 0.511 & $-17.91(14)$  & $ 11.00(15)$ & A \\
2013-09-02 & 6537.599166 & 0.982 & $12.23(15)$  & $-22.55(15)$ & A \\
2013-09-09 & 6544.591214 & 0.022 & $-0.18(25)$  & $-9.94(26)$ & A \\
2014-04-23 & 6770.897695 & 0.344 & $-25.70(15)$  & $ 17.52(15)$ & E \\
2014-05-17 & 6794.863326 & 0.484 & $-20.44(15)$  & $ 11.92(15)$ & E
\enddata
\label{table0}
\tablenotetext{a}{Exposure midpoint timestamp, (BJD--2450000)}
\tablenotetext{b}{T = TRES, A = ARCES, E = APOGEE}
\end{deluxetable}

\begin{figure*}[h!]
\begin{center}
\includegraphics[width=1.85\columnwidth]{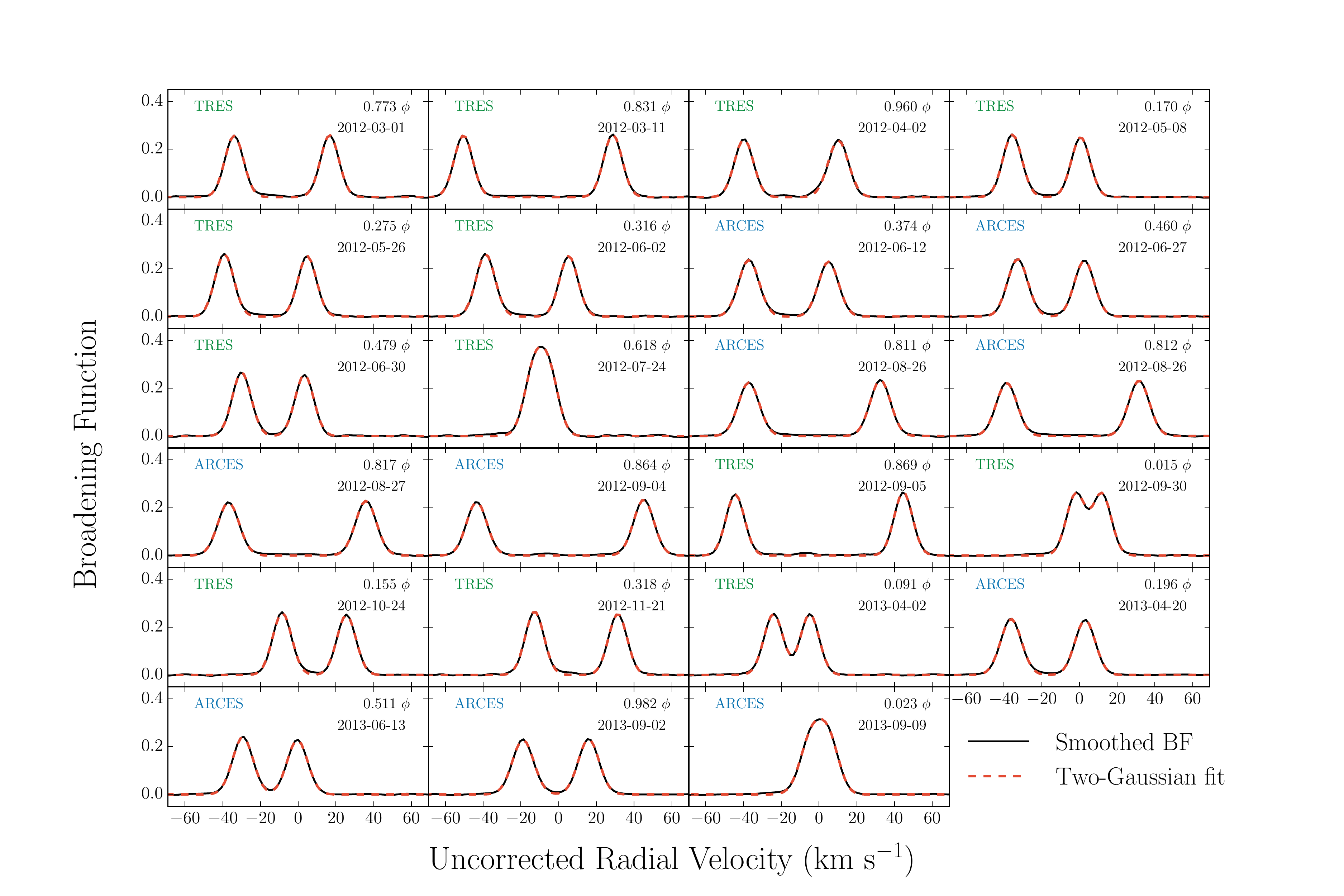}
\caption{\label{fig:bffig} Radial velocities extracted for 23 ARCES and TRES observations of KIC 9246715 with the broadening function (BF) technique. Each panel represents one spectral observation, ordered chronologically, for which the BF convolution of the target star with a template PHOENIX model spectrum is shown in black. To identify the location of each BF in radial velocity space, we fit a pair of Gaussians, which are plotted in red. The date of observation, orbital phase, and instrument used are printed in the upper corners of each panel. Barycentric corrections have not yet been applied to these velocities.
}
\end{center}
\end{figure*}

\begin{figure*}[h!]
\begin{center}
\includegraphics[width=1.6\columnwidth]{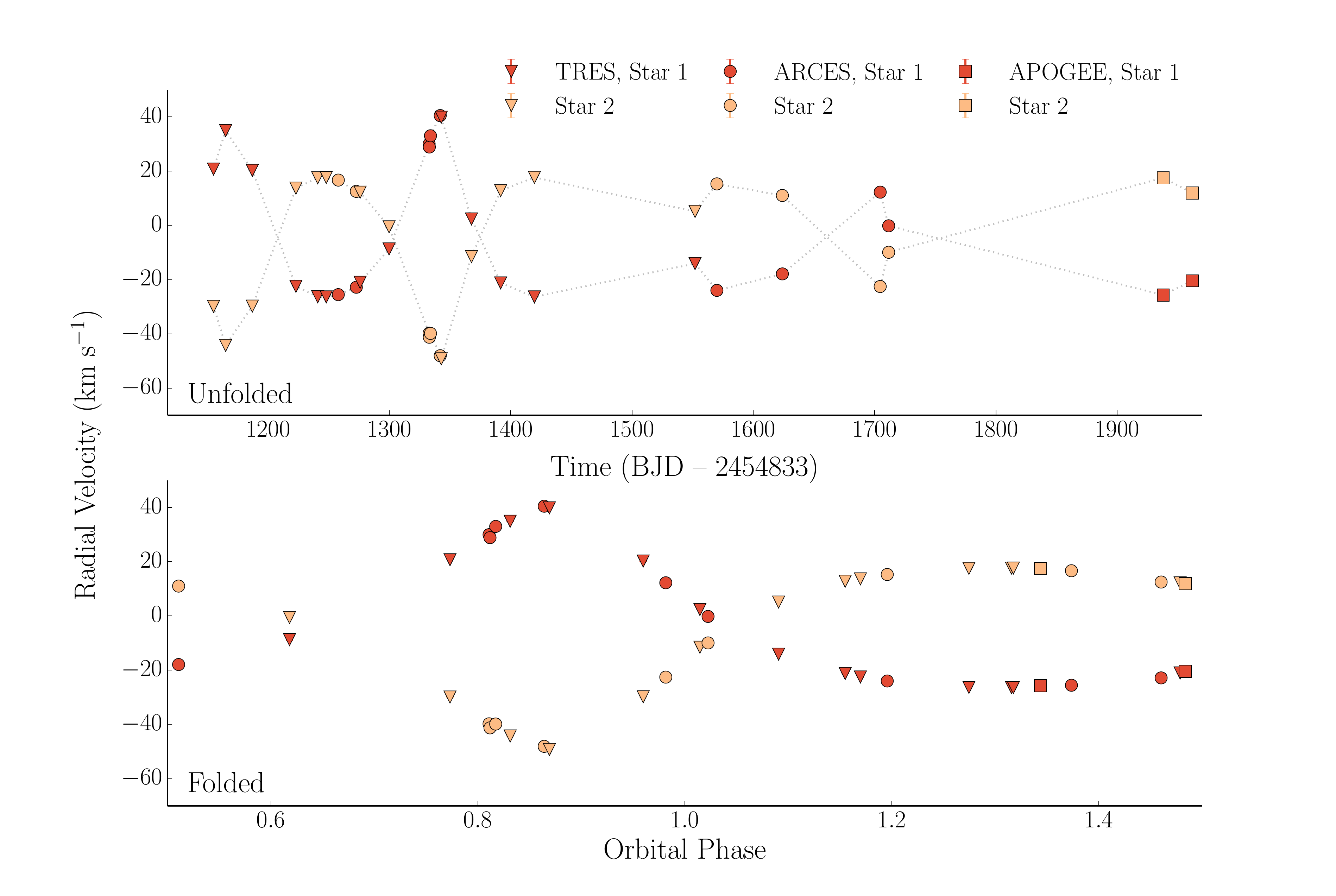}
\caption{\label{fig:rvfig} Radial velocity curves for both stars in KIC 9246715. The top panel shows the velocities as a function of time, with a light dotted line to guide the eye. The bottom panel shows the folded radial velocity curve over one orbit. Symbol shape indicates which spectrograph took each observation.
}
\end{center}
\end{figure*}

We use a PHOENIX BT-Settl model atmosphere spectrum as a BF template \citep{all03}. This particular model uses \citet{asp09} solar abundance values for a star with $T_{\rm{eff}} = 4800$ K, $\log g = 2.5$, and solar metallicity, selected based on revised KIC values for KIC 9246715\footnote{We later confirm that the RV results are indistinguishable from those measured with a more accurate BF model template ($T_{\rm{eff}} = 5000$ K, $\log g = 3.0$, see Table \ref{table2}).} \citep{hub14.2}. Since the BF handles line broadening between template and target robustly, we do not adjust the resolution of the template.

Using a model template avoids inconsistencies between the optical and IR regime, additional barycentric corrections, spurious telluric line peaks, and uncertainties from a template star's systemic RV. In comparison, we test the BF with an observation of Arcturus as a template, and find that using a real star template gives BF peaks that are narrower and have larger amplitudes. These qualities may be essential to measure RVs in the situation where a companion star is extremely faint, because the signal from a faint companion may not appear above the noise if the BF peaks are weaker and broader. However, each star contributes roughly equally to the overall spectrum here, so we choose a model atmosphere template for simplicity. The advantages of using a real star spectrum as a BF template instead of a model will likely be crucial for future work, as most other RG/EBs are composed of a bright RG and relatively faint main sequence companion.

For the optical spectra, we consider the wavelength range 5400--6700 \AA. This region is chosen because it has a high signal-to-noise ratio and minimal telluric features. For the near-IR APOGEE spectra, we consider the wavelength range 15150--16950 \AA. We smooth the BF with a Gaussian to remove un-correlated, small-scale noise below the size of the spectrograph slit, and then fit Gaussian profiles \revise{with a least-squares technique} to measure the location of the BF peaks in velocity space. The geocentric (uncorrected) results from the BF technique are shown for the optical spectra in Figure \ref{fig:bffig}. The results look similar for the near-IR spectra. The final derived radial velocity points with barycentric corrections are presented in Table \ref{table0} and Figure \ref{fig:rvfig}. The radial velocities vary from about $-50$ to $40 \ \rm{km} \ \rm{s}^{-1}$, \revise{with uncertainties on the order of $0.02 \ \rm{km} \ \rm{s}^{-1}$. Uncertainties are assigned based on the error in position from the least-squares best-fit Gaussian to each BF peak.}

\subsection{Comparison with TODCOR}\label{todcor}
To confirm that the BF-extracted radial velocities are accurate, we also use TODCOR \citep{zuc94} to extract radial velocities for the TRES spectra. TODCOR, which stands for two-dimensional cross-correlation, uses a template spectrum from a library with a narrow spectral range (5050--5350 \AA) to make a two-component radial velocity curve for spectroscopic binaries. It is commonly used with TRES spectra for eclipsing binary studies. From the radial velocity curve, TODCOR subsequently calculates an orbital solution. We use the full TODCOR RV extractor + orbital solution calculator for the TRES spectra, and compare this with the TODCOR orbital solution calculator for the combined ARCES, TRES, and APOGEE RV points which were extracted with the BF technique. We find that the two orbital solutions are in excellent agreement. The TODCOR RVs (available for TRES spectra only) are on average $0.22 \pm 0.25 \ \rm{km \ s}^{-1}$ systematically lower than the BF RVs, which we attribute to a physically unimportant difference in RV zeropoint.

\section{Stellar atmosphere model}\label{atm}

\subsection{Spectral disentangling}\label{disentangle}
Before the two stars' atmospheres can be modeled, it is necessary to extract each star's spectrum from the observed binary spectra. While the location of a set of absorption lines in wavelength space is the only requirement for radial velocity studies, using an atmosphere model to measure $T_{\rm{eff}}$, $\log g$, and metallicity [Fe/H] for each star requires precise equivalent widths of particular absorption lines.

To this end, we use the FDBinary tool \citep{ili04} on the spectral window 4900--7130 \AA \ to perform spectral decomposition. Following the approach in \citet{bec14}, we break the window into 222 pieces that each span about 10 \AA. FDBinary does not require a template, and instead uses the orbital parameters of a binary to separate a set of double-lined spectral observations in Fourier space. We test FDBinary's capabilities by creating a set of simulated double-lined spectra from a weighted sum of two identical spectra of Arcturus. When the orbital solution and flux ratio is correctly specified, the program returns a pair of single-lined spectra that are indistinguishable from the original.

FDBinary requires a set of double-lined spectral observations re-sampled evenly in $\ln \lambda$. For each input spectrum, it is important to apply barycentric corrections and subtract the binary's systemic velocity ($-4.48$ km s$^{-1}$ in this case, see Section \ref{model} and Table \ref{table1}). FDBinary further requires six parameters to define the shape of the radial velocity curve: orbital period, time of periastron passage (zeropoint), eccentricity, longitude of periastron, and amplitudes of each star's radial velocity curve. We set these to 171.277 days, 319.7 days\footnote{Units of BJD--2454833}, 0.35, 17.3 deg, 33.1 km s$^{-1}$, and 33.4 km s$^{-1}$, respectively. While FDBinary does include an optimization algorithm for any subset of these parameters, we use more robust fixed values from a \revise{preliminary dynamical model similar to the ones in} Section \ref{model}. Finally, FDBinary requires a light ratio for each observation. Because the two stars are so similar, and none of our spectra were taken during eclipse, we set all light ratios to 1. This is further justified by the nearly-equal amplitude of each star's broadening function (see Figure \ref{fig:bffig}). We tried adjusting the light ratio and found that the result is qualitatively similar, but systematically increases the strength of all features in one spectrum while systematically decreasing the strength of all features in the other.

\begin{figure*}[ht!]
\begin{center}
\includegraphics[width=2.1\columnwidth]{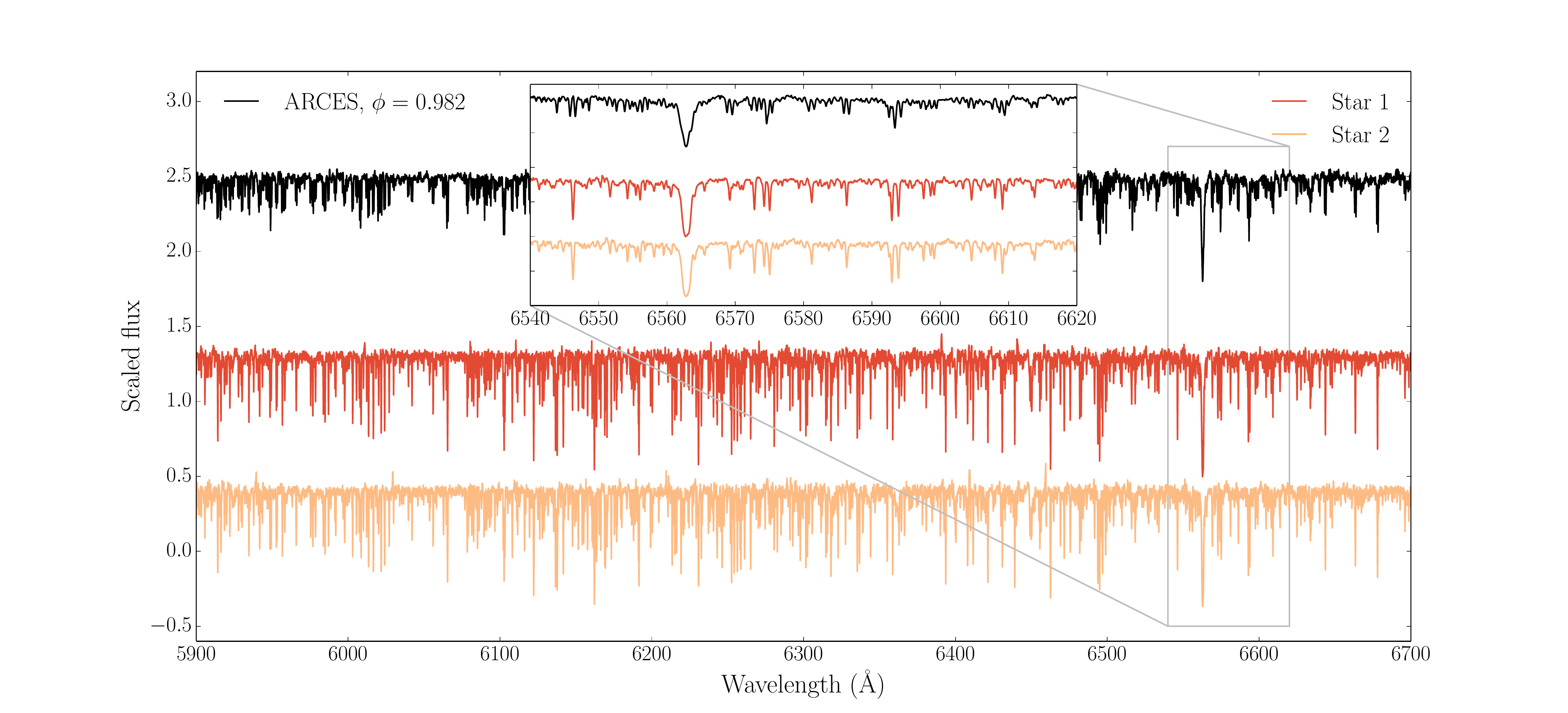}
\caption{\label{fig:twospectra} Disentangled spectra from FDBinary for the two stars in KIC 9246715. The y-axis is offset by an arbitrary amount for clarity. For comparison, a typical observation from the ARCES spectrograph taken close to primary eclipse ($\phi = 0.982$) on 2013-09-02 is in black. The zoom panel is a clearer view of individual spectral features, including $\textrm{H}\alpha$, and clearly shows that the observed double-lined spectrum has been decomposed into two single-lined components. The full decomposed spectra span 4900--7130 \AA; only a portion is shown here.
}
\end{center}
\end{figure*}

All 23 optical spectra of KIC 9246715 are processed together in FDBinary, and the result is a pair of disentangled spectra with zero radial velocity. A portion of the resulting individual spectra are shown in Figure \ref{fig:twospectra} with a characteristic ARCES spectrum containing signals from both stars for comparison.

\subsection{Parameters from atmosphere modeling}\label{parameters}
We use the radiative transfer code MOOG \citep{sne73} to estimate $T_{\rm{eff}}$, $\log g$, and metallicity [Fe/H] for the disentangled spectrum of each star in KIC 9246715. First, we use ARES (Automatic Routine for line Equivalent widths in stellar Spectra, \citealt{Sousa_2007}) with a modified {\rm Fe}\kern 0.1em{\sc i} and {\rm Fe}\kern 0.1em{\sc ii} linelist from \citet{tsa13}. ARES automatically measures equivalent widths for spectral lines which can then be used by MOOG. An excellent outline of the process is given by \citet{Sousa_2014}.

We use ARES to identify 66 {\rm Fe}\kern 0.1em{\sc i} and 9 {\rm Fe}\kern 0.1em{\sc ii} lines in the spectrum of Star 1, and 74 {\rm Fe}\kern 0.1em{\sc i} and 10 {\rm Fe}\kern 0.1em{\sc ii} lines in the spectrum of Star 2, all in the 4900--7130 \AA \ region. To arrive at a best-fit stellar atmosphere model with MOOG, we follow the approach of \citet{mag13}. Error bars are determined based on the standard deviation of the derived abundances and the range spanned in excitation potential or equivalent width. For Star 1, we find $T_{\rm{eff}} = 4990 \pm 90 \ \rm{K}$, $\log g = 3.21 \pm 0.45$, and $\rm{[Fe/H]} = -0.22 \pm 0.12$, with a microturbulence velocity of $1.86 \pm 0.09 \ \rm{km \ s}^{-1}$. For Star 2, we find $T_{\rm{eff}} = 5030 \pm 80 \ \rm{K}$, $\log g = 3.33 \pm 0.37$, and $\rm{[Fe/H]} = -0.10 \pm 0.09$, with a microturbulence velocity of $1.44 \pm 0.09 \ \rm{km \ s}^{-1}$.

Projected rotational velocities can also be measured from stellar spectra. To estimate this, we compare the disentangled spectra to a grid of rotationally broadened spectra. We find both stars have $v_{\rm{broad}} \simeq 8 \ \rm{km \ s}^{-1}$. It is important to consider that this observed broadening is a combination of each star's rotational velocity and macroturbulence: $v_{\rm{broad}} = v_{\rm{rot}} \sin i + \zeta_{\rm{RT}}$, where $\zeta_{\rm{RT}}$ is the radial-tangential macroturbulence dispersion \citep{gra78}. We note that rotational broadening is Gaussian while broadening due to macroturbulence is cuspier, but these subtle line profile differences are not distinguishable here. \citet{car08} find a large range of macroturbulence dispersions for giant stars which may vary as a function of luminosity, gravity, and temperature, and introduce a non-physically-motivated empirical relation $v_{\rm{broad}} = [(v_{\rm{rot}} \sin i)^2 + 0.95~\zeta_{\rm{RT}}^2]^{1/2}$, while \citet{tay15} estimate the macroturbulence for giant stars to be on order $10 \%$ of the observed broadening. In any case, at least some of the observed line broadening is attributable to macroturbulence, and we conclude neither star in KIC 9246715 is a particularly fast rotator.

\section{Physical parameters from light curve \& radial velocities}\label{model}
To derive physical and orbital parameters for KIC 9246715, we use the Eclipsing Light Curve (ELC) code \citep{oro00}. ELC computes model light and velocity curves and uses \revise{a Differential Evolution Markov Chain Monte Carlo optimizing algorithm \citep{ter06}} to simultaneously solve for a suite of stellar parameters. It is able to consider any set of input constraints simultaneously, i.e., a combination of light curves and radial velocities, and can use a full treatment of Roche geometry \citep{kop69,avn75}. ELC uses a grid of NextGen model atmospheres integrated over the \emph{Kepler} bandpass to assign an intensity at the surface normal of each star. Intensities for the other portions of each star's visible surface are then computed with a quadratic limb darkening law. By including the temperature of Star 1 as a fit parameter, ELC will try different model atmospheres, thereby indirectly computing stellar temperature. ELC uses $\chi^2$ as a measure of fitness to refine a best-fit model:

\begin{eqnarray}
\chi^2 & = &
\sum_i \frac{ (f_{\rm{mod}}(\phi_i; \ {\bf a}) - f_{\rm{obs}}(\rm{\it{Kepler}}))^2 }{\sigma_i^2(\rm{\it{Kepler}})} \nonumber \\
& + & \sum_i \frac{ (f_{\rm{mod}}(\phi_i; \ {\bf a}) - f_{\rm{obs}}(\rm{RV_1}))^2 }{\sigma_i^2(\rm{RV_1})} \\
& + & \sum_i \frac{(f_{\rm{mod}}(\phi_i; \ {\bf a}) - f_{\rm{obs}}(\rm{RV_2}))^2}{\sigma_i^2(\rm{RV_2})}, \nonumber
\end{eqnarray}

\noindent where $f_{\rm{mod}}(\phi_i; \ {\bf a})$ is the ELC model flux at a given phase $\phi_i$ for a set of parameters ${\bf a}$, $f_{\rm{obs}}$ is the observed value at the same phase, and $\sigma_i$ is the associated uncertainty.

We compute two sets of ELC models: the first uses all eclipses from the light curve together with all radial velocity points, and the second breaks the light curves into segments to investigate how photometric variations from one orbit to another affect the results. Both sets of models employ ELC's ``fast analytic mode.'' This uses the equations in \citet{man02} to treat both stars as spheres, which is reasonable for a well-detached binary like KIC 9246715 ($R/a < 0.04$ for both stars). \revise{The results from both sets of models are presented in Table \ref{table1}. We adopt the ``All-eclipse model'' as the accepted solution, for reasons described below.}

\begin{figure*}[ht!]
\begin{center}
\includegraphics[width=1.95\columnwidth]{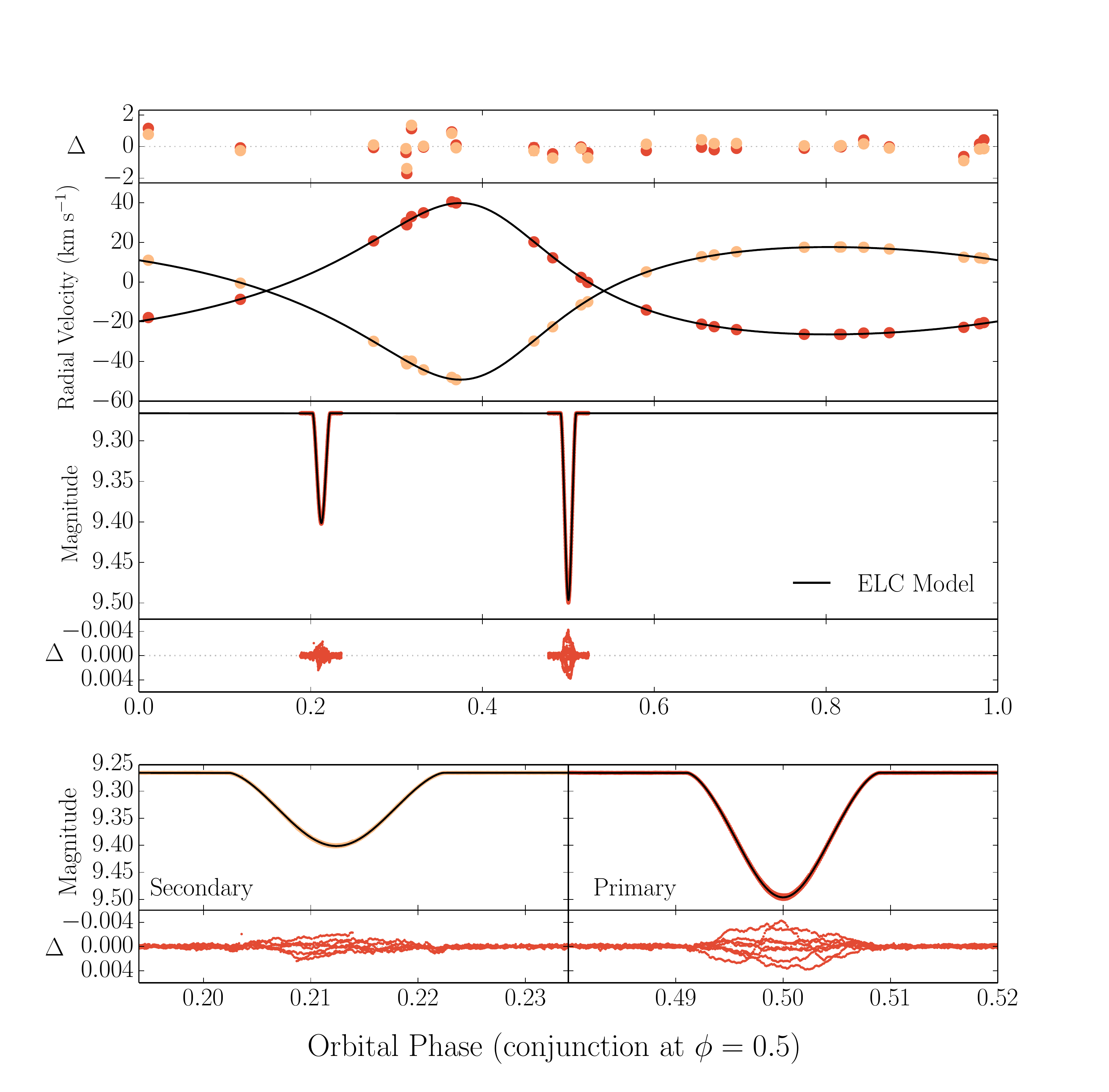}
\caption{\label{fig:ELCresult} ELC model for all eclipses of KIC 9246715 taken together. The top two panels show the folded radial velocities, while the middle two panels show the folded light curve. A single full orbit is shown. The bottom four panels are a zoom of each eclipse. Residuals are indicated by a $\Delta$ symbol. Red and yellow points are observations and the black line is the all-eclipse ELC model fit. The primary and secondary eclipses are the same configurations as illustrated in Figure \ref{fig:keplerfig}. \newrevise{While one primary eclipse epoch suffers from increased contamination due to a nearby star (see Section \ref{segment}), the overall scatter in the eclipse residuals is greater during primary eclipse than during secondary eclipse. This suggests Star 1 is more active than Star 2, and is discussed further in Section \ref{actrot}.}
}
\end{center}
\end{figure*}

\begin{deluxetable*}{lccc}
\tablecolumns{4}
\tablewidth{0pt}
\tabletypesize{\footnotesize}
\tablecaption{Physical parameters of KIC 9246715 from ELC modeling.}
\centering
\tablehead{
\colhead{Parameter}	& \colhead{All-eclipse model}		& \colhead{LC segment RMS}		& \colhead{Comment}	
}
\startdata
$P_{\rm{orb}} \ \rm{[d]}$	&	$171.27688 \pm 0.00001$	& 	$171.276 \pm 0.001$		& 	 \\[0.4em]
$T_{\rm{conj}}\ \rm{[d]}$	&	$ 337.51644 \pm 0.00005$		& 	$ 337.519 \pm 0.002$	 	& 	0 d $\equiv$ 2454833 BJD	\\[0.4em]
$i$ [deg]			&	$87.051\substack{+0.009 \\ -0.003}$			& 	$87.08 \pm 0.03 $			&	 \\[0.4em]
$e$				&	$0.3559\substack{+0.0002 \\ -0.0003}$		&	$0.355 \pm 0.001$		&	 \\[0.4em]
$\omega$ [deg]		&	$18.4\substack{+0.1 \\ -0.2}$			&	$17.7 \pm 0.7$			&	 \\[0.4em]
$e \cos \omega$	&	$0.33773\substack{+0.00005 \\ -0.00003}$		& 	$0.3379 \pm 0.0001$ 		&	 \\[0.4em]
$e \sin \omega$	&	$0.1123\substack{+0.0007 \\ -0.0012}$		& 	$0.108 \pm 0.004$ 		&	 \\[0.4em]
$T_2/T_1$		&	$1.001\substack{+0.001 \\ -0.002}$		& 	$0.993 \pm 0.008$ 		&	 \\[0.4em]
$a \ [R_{\odot}]$	&	$211.3\substack{+0.2 \\ -0.3}$				& 	$211.0 \pm 0.3$		 	&	 \\[0.4em]
contam			&	$0.002\substack{+0.004 \\ -0.001}$			& 	$0.02 \pm 0.01$ 			&	\emph{Kepler} contamination \\[0.4em]
$\gamma_{\rm{vel}}$ [km s$^{-1}$]	& $-4.4779 \pm 0.002$ 	& 	$-4.4797 \pm 0.0007$			&	systemic velocity$^{\rm{a}}$ \\
\cutinhead{Star 1}
$M \ [M_{\odot}]$	&	$2.171\substack{+0.006 \\ -0.008}$			& 	$2.162 \pm 0.008$ 		&	 \\[0.4em]
$R \ [R_{\odot}]$	&	$8.37\substack{+0.03 \\ -0.07}$			& 	$8.27 \pm 0.09$			&	 \\[0.4em]
$R/a$			&	$0.0396\substack{+0.0001 \\ -0.0003}$		& 	$0.0392 \pm 0.0004$ 		&	 \\[0.4em]
$T$ [K]			&	$4930\substack{+140 \\ -230}$				& 	\nodata			&	 \\[0.4em]
$K$ [km s$^{-1}$]	&	$33.19\substack{+0.04 \\ -0.05}$				& 	$33.13 \pm 0.06$		  	&	 \\[0.4em]
$\log g$ [cgs]		&	$2.929\substack{+0.007 \\ -0.003}$			& 	$2.938 \pm 0.008$ 			&	 \\[0.4em]
$q_1$			&	$0.66\substack{+0.02 \\ -0.04}$			& 	$0.72 \pm 0.02$ 				&	triangular limb darkening$^{\rm{b}}$ \\[0.4em]
$q_2$			&	$0.25\substack{+0.02 \\ -0.01}$			& 	$0.31 \pm 0.02$ 			&	triangular limb darkening$^{\rm{b}}$ \\
\cutinhead{Star 2}
$M \ [M_{\odot}]$	&	$2.149\substack{+0.006 \\ -0.008}$			& 	$2.140 \pm 0.008$ 			&	 \\[0.4em]
$R \ [R_{\odot}]$	&	$8.30\substack{+0.04 \\ -0.03}$			& 	$8.29 \pm 0.01$		 	&	 \\[0.4em]
$R/a$			&	$0.0393 \pm 0.0001$		& 	$0.03928 \pm 0.00002$ 		&	 \\[0.4em]
$T$ [K]			&	$4930\substack{+140 \\ -230}$				& 	\nodata 			&	 \\[0.4em]
$K$ [km s$^{-1}$]	&	$33.53\substack{+0.04 \\ -0.05}$				& 	$33.47 \pm 0.06$ 			&	 \\[0.4em]
$\log g$ [cgs]		&	$2.932\substack{+0.003 \\ -0.004}$			& 	$2.9315 \pm 0.0005$ 			&	 \\[0.4em]
$q_1$			&	$0.55\substack{+0.03 \\ -0.04}$			& 	$0.52 \pm 0.05$ 			&	triangular limb darkening$^{\rm{b}}$ \\[0.4em]
$q_2$			&	$0.33 \pm 0.02$			& 	$0.41 \pm 0.02$ 			&	triangular limb darkening$^{\rm{b}}$
\enddata
\label{table1}
\tablenotetext{a}{The uncertainties reported for $\gamma_{\rm{vel}}$ are based on the internal consistency of the model using relative velocities. The true error is on the order of 0.2--0.3 km s$^{-1}$ (Section \ref{todcor}).}
\tablenotetext{b}{The triangular limb darkening law (Kipping 2013) re-parameterizes the quadratic limb darkening law, $I(\mu)/I(1) = 1 - u_1(1 - \mu) - u_2(1 - \mu)^2$, with new coefficients $q_1 \equiv (u_1 + u_2)^2$ and $q_2 \equiv 0.5 u_1 (u_1 + u_2)^{-1}$.}
\end{deluxetable*}

\subsection{All-eclipse ELC model}\label{all-eclipse}
We use ELC \revise{to compute more than 2 million models which fit 16 parameters}: orbital period $P_{orb}$, zeropoint $T_{conj}$ (this sets the primary eclipse to orbital phase $\phi_{ELC} = 0.5$ instead of $\phi = 0$), orbital inclination $i$, $e \sin \omega$ and $e \cos \omega$ (where $e$ is eccentricity and $\omega$ is the longitude of periastron), the temperature of the primary star $T_1$, the mass of the primary star $M_1$, the amplitude of the primary star's radial velocity curve $K_1$, the fractional radii of each star $R_1/a$ and $R_2/a$, the temperature ratio $T_2/T_1$, the \emph{Kepler} contamination factor, and stellar limb darkening parameters for the triangular limb darkening law \citep{kip13}. The scale of the system (and hence the component masses and radii) is uniquely determined given the primary star mass, the amplitude of its radial velocity curve, and the orbital period. \newrevise{Error bars are determined from the cumulative distribution frequency of each fit parameter after the first 10,000 models are excluded to allow for an appropriate MCMC burn-in period. Quoted values are 50\% of the cumulative distribution function with the one-sigma upper error at 84.25\% and one-sigma lower error at 15.75\%.} The results are in Figure \ref{fig:ELCresult} and Table \ref{table1}.

\subsection{Light curve segment ELC models}\label{segment}
To investigate secular changes in KIC 9246715, we split the \emph{Kepler} light curve into seven segments such that each contains one primary and one secondary eclipse. This is particularly motivated by \revise{the photometric variability seen in Figure \ref{fig:lcfig2} and} the residuals of the primary eclipse in the all-eclipse model, as shown in Figure \ref{fig:ELCresult}. Of all the observed primary eclipses, the one in the seventh light curve segment is slightly shallower than the others by about 0.004 magnitudes. \revise{To learn why, we examine the \emph{Kepler} Target Pixel Files, which reveal the aperture used for KIC 9246715 includes a larger portion of a nearby contaminating star every fourth quarter. This higher contamination is coincident with the secondary eclipse in the fifth light curve segment and both eclipses in the seventh light curve segment. Higher contamination results in shallower eclipses because there is an overall increase in flux, and we conclude that the shallower primary eclipse is a result of this contamination rather than a star spot or other astrophysical signal.}

We therefore calculate a second set of parameters based on the root-mean-square (RMS) of \revise{six ELC models, one for each light curve segment, excluding the seventh segment which has significantly higher contamination in both eclipses}. Each segment still includes the full set of radial velocity data. The values reported are the RMS of these seven models, $a_{\rm{RMS}} = \sqrt{\frac{1}{n} \sum_{i=1}^n (a_i^2)}$, plus or minus the RMS error, $\sqrt{\frac{1}{n} \sum_{i=1}^n (a_i - a_{\rm{RMS}})^2}$. These are reported in Table \ref{table1}. \revise{Temperature is not reported because the white-light \emph{Kepler} bandpass is not well-suited to constrain stellar temperatures, and the RMS errors among the light curve segments are artificially small.}

For all parameters, the all-eclipse model and the LC segment model agree \revise{within $2\sigma$. We note that $\omega$, the \emph{Kepler} contamination, and $R_1$ all have significantly larger uncertainties in the LC segment results than the all-eclipse results. This reflects an inherent degeneracy between viewing angle and stellar radius in a binary with grazing eclipses, which is exacerbated by uncertainties in limb darkening and temperature, as well as varying contamination between quarters. When we hold both stars' limb darkening coefficients fixed with theoretical values $q_1 = 0.49$ and $q_2 = 0.37$ \citep{cla13}, we find an ELC solution that gives $R_1 \simeq 7.9 \ R_\odot$, $R_2 \simeq 8.2 \ R_\odot$, $\omega \simeq 17.4 \ \rm{deg}$, and contamination as high as 5 \%. However, this solution has a higher $\chi^2$ than the models which allow triangularly sampled quadratic limb darkening coefficients \citep{kip13} to be free parameters, and it is important to consider that theoretical limb darkening values are poorly constrained for both giant stars and wide bandpasses. We therefore adopt the all-eclipse ELC solution in this work because it has the lowest $\chi^2$ and uses all available data to constrain the system.}

\section{Discussion}\label{discuss}

\subsection{Comparison with asteroseismology}\label{seismo}
We expect both evolved giants in KIC 9246715 to exhibit solar-like oscillations. These should be observable as pure p-modes for radial oscillations ($\ell = 0$), mixed p- and g- modes for dipolar oscillations ($\ell = 1$), and p-dominated modes for quadrupolar oscillations ($\ell = 2$) in \emph{Kepler} long-cadence data. For solar-like oscillators, the average large frequency separation between consecutive p-modes of the same spherical degree $\ell$, $\Delta \nu$, has been shown to scale with the square root of the mean density of the star. The frequency of maximum oscillation power, $\nu_{\rm{max}}$, carries information about the physical conditions near the stellar surface and is a function of surface gravity and effective temperature \citep{kje95}. These scaling relations may be used to estimate a star's mean density and surface gravity:

\begin{equation} \label{density}
{\frac{\bar{\rho}}{\bar{\rho}_{\odot}}} \simeq {\left( \frac{\Delta \nu}{\Delta \nu_{\odot}} \right)}^{2}
\end{equation}

\noindent and

\begin{equation} \label{gravity}
{\frac{g}{g_{\odot}}} \simeq {\left( \frac{\nu_{\rm{max}}}{\nu_{\rm{max}, \ \odot}} \right)} {\left( \frac{T_{\rm{eff}}}{T_{\rm{eff}, \ \odot}} \right)}^{-1/2}.
\end{equation}

\revise{Equation} \ref{density} is valid only for oscillation modes of large radial order $n$, where pressure modes can be mathematically described in the frame of the asymptotic development \citep{tas80}. Even though red giants do not perfectly match these conditions, because the observed oscillation modes have \revise{small radial orders on the order of $n \sim 10$}, the scaling relations do appear to work. Quantifying how well they work and in what conditions is more challenging. This is why measuring oscillating stars' masses and radii independently from seismology is so important.

Surprisingly, when \citet{gau13} and \citet{gau14} analyzed the oscillation modes of KIC 9246715 to estimate global asteroseismic parameters, only one set of modes corresponding to a single oscillating star was found. Of the 18 oscillating RG/EBs in the \emph{Kepler} field, KIC 9246715 is the only one with a pair of giant stars (the rest are composed of a giant star and a main sequence star).

In addition, the light curve displays photometric variability as large as 2\% peak-to-peak, \revise{as shown in Figure \ref{fig:lcfig2}}, which is typical of the signal created by spots on stellar surfaces. The pseudo-period of this variability was observed to be about half the orbital period, which suggests resonances in the system. \citet{gau14} speculated that star spots may be responsible for inhibiting oscillations on the smaller star, and a similar behavior was observed in other RG/EB systems. In this section, we reestimate the global seismic parameters of the oscillation spectrum that was previously identified (Section \ref{subsubsec_main_osc}), \revise{analyze the mixed oscillation modes to determine the oscillating star's evolutionary state (Section \ref{subsubsec_mixed}), \newrevise{investigate} which star is more likely to be exhibiting oscillations (Section \ref{identifying}), and address the discrepancy between different surface gravity measurements (Section \ref{gravity_compare}).}

\subsubsection{Global asteroseismic parameters of the oscillating star}
\label{subsubsec_main_osc}
We now re-estimate $\nu_{\rm{max}}$ and $\Delta \nu$ for the oscillation spectrum in the same way as \citet{gau14}, but by using the whole \textit{Kepler} dataset (Q0--Q17). \revise{The frequency at maximum amplitude of solar-like oscillations $\nu_{\rm{max}}$ is measured by fitting the mode envelope with a Gaussian function and the background stellar activity with a sum of two semi-Lorentzians. The large frequency separation $\Delta\nu$ is obtained from the filtered autocorrelation of the time series \citep{mos09}.} Differences with respect to previous estimates are negligible, as we find $\nu_{\rm{max}} = 106.4 \pm 0.8$ and $\Delta \nu = 8.31 \pm 0.02 \ \mu \rm{Hz}$. Because the ELC results yield $T_2/T_1=0.989$ (Table \ref{table1}) and the stellar atmosphere analysis gives $T_1 = 4990 \pm 90 \ \rm{K}$ and $T_2 = 5030 \pm 80 \ \rm{K}$ (Section \ref{parameters}), we use an effective temperature of $T_{\rm{eff}} = 5000 \pm 100 \ \rm{K}$ in the asteroseismic scaling equations. \newrevise{Assuming a single oscillating star, the mode amplitudes are only $\sim 60\%$ as high as expected ($A_{\rm{max}}(\ell=0) \simeq 15$ ppm, and not 6.6 ppm as erroneously reported by \citealt{gau14}) when compared to the $\sim 24$ ppm predicted from mode amplitude scaling relations \citep{cor13}. The modes are four times wider than expected as well, with $\ell=0$ linewidths $\simeq 0.4 \ \mu \rm{Hz}$ near $\nu_{\rm{max}}$ rather than a value closer to $0.1 \ \mu \rm{Hz}$ as predicted for stars with similar $\nu_{\rm{max}}, \Delta \nu,$ and $T_{\rm{eff}}$ \citep{cor15}.}

To determine mass, radius, surface gravity, and mean density, we use the scaling relations after correcting $\Delta \nu$ for the red giant regime \citep{mos13}\footnote{\newrevise{Other scaling relation applications, such as \citet{cha11} and \citet{kal10}, assume the observed $\Delta \nu$ is equal to the asymptotic $\Delta \nu$. \citet{mos13} uses a correction factor to account for the fact that oscillating red giants are not in the asymptotic regime, which we apply here.}}. In essence, instead of directly plugging the observed $\Delta \nu_{\rm{obs}}$ into Equations \ref{density} and \ref{gravity}, we estimate the asymptotic large spacing via $\Delta \nu_{\rm{as}} = \Delta \nu_{\rm{obs}} (1 + \zeta)$, where $\zeta = 0.038$. With this correction of the large spacing, we obtain $M = 2.17 \pm 0.14 \ M_{\odot}$ and $R = 8.26 \pm 0.18 \ R_{\odot}$. In terms of mean density and surface gravity, which independently test the $\Delta \nu$  and $\nu_{\rm{max}}$ relations, respectively, we find $\bar{\rho}/\bar{\rho}_{\odot} = (3.86 \pm 0.02) \times 10 ^{-3}$ and $\log g = 2.942 \pm 0.008$. A comparison of key parameters determined from all our different modeling techniques is in Table \ref{table2}.

\begin{deluxetable*}{lccccc}
\tablecolumns{6}
\tablewidth{0pt}
\tabletypesize{\footnotesize}
\tablecaption{Physical parameter comparisons for KIC 9246715 with different modeling techniques.}
\centering
\tablehead{
\colhead{} \vspace{-0.15cm} 		& \colhead{Mass} 		& \colhead{Radius} 		& \colhead{$\log g$} & \colhead{$\bar{\rho}$}	& \colhead{$T_{\rm{eff}}$}	 \\
\colhead{Model}  \vspace{-0.15cm} 	& \colhead{} 			& \colhead{} 			& \colhead{} 		& \colhead{}		& \colhead{} 			\\
\colhead{} 					& \colhead{($M_{\odot}$)} & \colhead{($R_{\odot}$)} & \colhead{(cgs)}	& \colhead{($\bar{\rho}_{\odot} \times 10^{-3}$)} & \colhead{(K)} 
}
\startdata
ELC (Light Curve + RV), Star 1		& $2.171\substack{+0.006 \\ -0.008}$	& $8.37\substack{+0.03 \\ -0.07}$	& $2.929\substack{+0.007 \\ -0.003}$		& $3.70\substack{+0.04 \\ -0.09}$		& $4930\substack{+140 \\ -230}$	\\[0.4em]
ELC (Light Curve + RV), Star 2		& $2.149\substack{+0.006 \\ -0.008}$	& $8.30\substack{+0.04 \\ -0.03}$	& $2.932\substack{+0.003 \\ -0.004}$		& $3.76\substack{+0.06 \\ -0.04}$ 	& $4930\substack{+140 \\ -230}$	\\[0.4em]
MOOG Stellar Atmosphere, Star 1 		& \nodata			& \nodata	 		& $3.21 \pm 0.45$	& \nodata	& $4990 \pm 90$	\\[0.4em]
MOOG Stellar Atmosphere, Star 2 		& \nodata			& \nodata	 		& $3.33 \pm 0.37$	& \nodata	& $5030 \pm 80$	\\[0.4em]
Global Asteroseismology, \revise{Star 1}\tablenotemark{a}		& \nodata	& \nodata	& \nodata		& $4.14 \pm 0.02$ & \nodata \\[0.4em]
Global Asteroseismology, \revise{Star 2}\tablenotemark{a}		& $2.17 \pm 0.14$	& $8.26 \pm 0.18$	& $2.942 \pm 0.008$		& $3.86 \pm 0.02$ & \tablenotemark{b}
\enddata
\label{table2}
\tablenotetext{a}{\revise{As discussed in Sections \ref{identifying} and \ref{subsec_second_osc}, we tentatively assign Star 2 to the main set of oscillations and Star 1 to the marginally detected oscillations.}}
\tablenotetext{b}{A fixed temperature of $5000 \pm 100 \ \rm{K}$ was assumed to calculate the other asteroseismic parameters.}
\end{deluxetable*}

\begin{figure*}[ht!]
\begin{center}
\includegraphics[width=2.1\columnwidth]{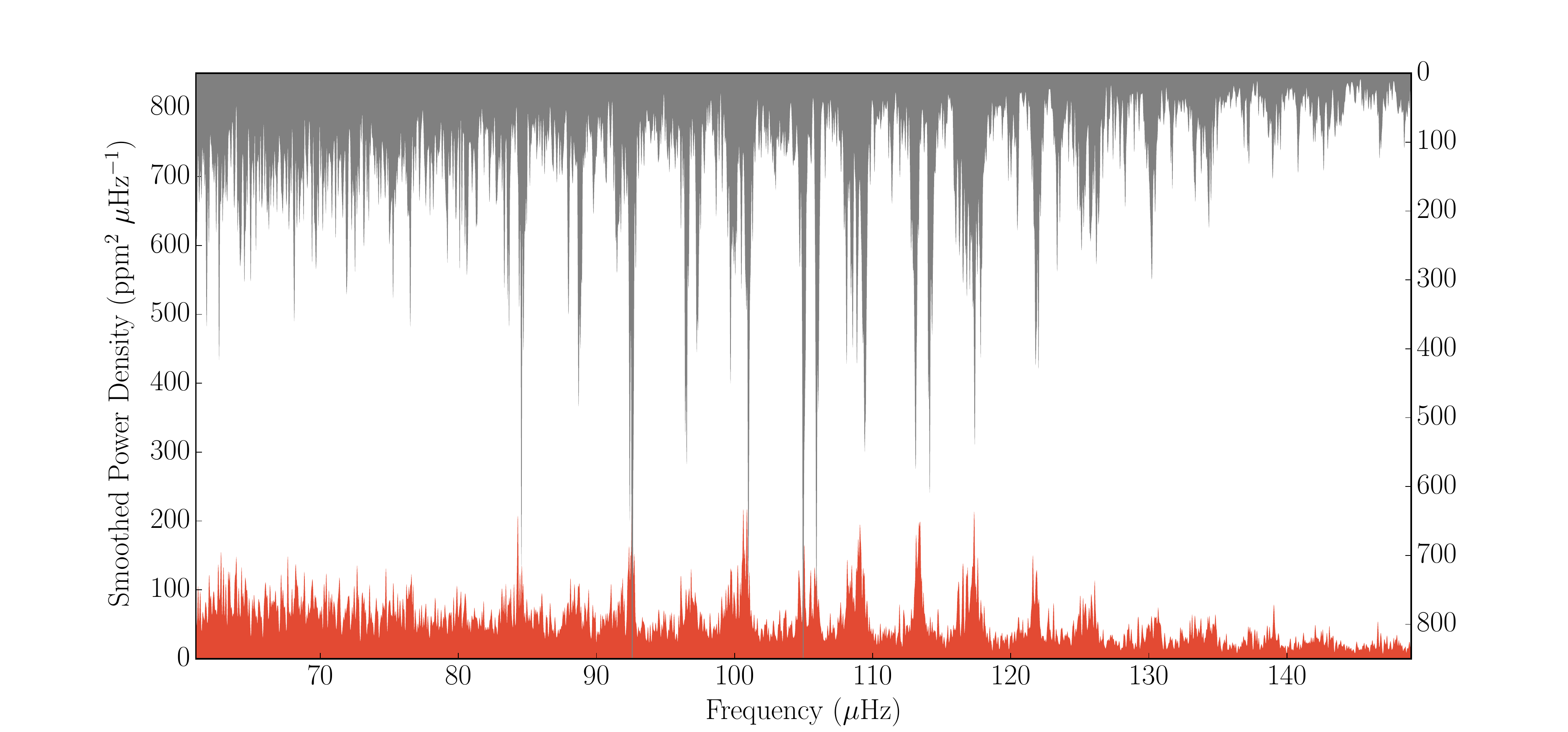}
\caption{\label{fig:twin}
Power density spectrum of KIC 11725564 (gray), a seismic ``twin'' of KIC 9246715 (red). Both power spectra are smoothed with a boxcar of $1/50$ of the large separation. The modes in KIC 11725564, a secondary red clump star, have $\Delta \nu$ and $\nu_{\rm{max}}$ which very nearly match KIC 9246715. They are less noisy and have larger amplitudes than the modes in KIC 9246715, making this star a useful asteroseismic comparison.
}
\end{center}
\end{figure*}

\subsubsection{Mixed oscillation modes}
\label{subsubsec_mixed}
Based on the distribution of mixed $\ell = 1$ modes, \citet{gau14} reported that the oscillation pattern period spacing was typical of that of a star from the secondary red clump, i.e., a core-He-burning star that has not experienced a helium flash. This was based on a \revise{dipole gravity mode} period spacing of $\Delta \Pi_1 \simeq 150 \ \rm{s}$. Red giant branch stars have smaller period spacings than red clump stars, and ($\Delta \Pi_1 = 150 \ \rm{s}$, $\Delta \nu = 8.31 \ \mu \rm{Hz}$) puts the oscillating star on the very edge of the asteroseismic parameter space that defines the secondary red clump \citep{mos14}. \revise{Due to noise and damped oscillations, it is difficult to unambiguously determine the mixed mode pattern described by \citet{mos12}. To more accurately assess the evolutionary stage of the oscillating star in KIC 9246715, we employ three different techniques to identify and characterize mixed modes.}

\revise{First, we perform a Bayesian fit to the individual oscillation modes of the star using the \textsc{D\large{iamonds}} code \citep{cor14} and the methodology for the peak bagging analysis of a red giant star in \citet{cor15}. We then compare the set of the obtained frequencies of mixed dipole modes with those from the asymptotic relation proposed by \citet{mos12}, which we compute using different values of $\Delta \Pi_1$. The result shows a significantly better match when values of $\Delta \Pi_1$ around $200 \ \rm{s}$ are used. This confirms that the oscillating star is settled on the core-He-burning phase of stellar evolution.} \newrevise{The results of the \textsc{D\large{iamonds}} fit are in the Appendix.}

\revise{Second, we search for stars with a power density spectrum that resembles the oscillation spectrum of KIC 9246715. As shown in Figure \ref{fig:twin}, a good match is found with the star KIC 11725564, which exhibits very similar radial and quadrupole modes as well as the mixed mode pattern. To find this ``twin,'' we calculate the autocorrelation of the KIC 9246715 oscillation spectrum, pre-whiten its radial and quadrupole modes, and convert it into period. We find a weak, broad peak at about $\Delta P_{\rm{obs}} = 80 \ \rm{s}$. A similar result of $\Delta P_{\rm{obs}} = 87 \ \rm{s}$ is found for KIC 11725564, with a notably cleaner signal thanks to higher mode amplitudes. This corresponds to the observed period spacing as defined by \citet{bed11} and \citet{mos11}, and indicates that the star is indeed likely to be a secondary clump star.}

\revise{Finally, we measure the asymptotic period spacing with the new method developed by \citet{mos15}. The signature $\Delta \Pi_1 = 150.4 \pm 1.4 \ \rm{s}$ is very clear, despite binarity. In fact, the presence of a second oscillation spectrum cannot mimic a mixed-mode pattern because its global amplitude is too small for us to observe a mode disturbance. Only one signature of an oscillating star is visible in a period spacing diagram.}

\revise{We conclude that the mixed oscillation modes in KIC 9246715 are indicative of a secondary red clump star.} This result is supported statistically by \citet{mig14}, who report it is more likely to find red clump stars than red giant branch stars in asteroseismic binaries in \emph{Kepler} data. This is largely due to the fact that evolved stars spend more time on the horizontal branch than the red giant branch. \newrevise{Due to the large noise level of the mixed modes, we are unable to measure a core rotation rate in the manner of \citet{bec12} and \citet{mos12}. However, the mixed modes appear to be doublets which support an inclination near 90 degrees.}

\subsubsection{Identifying the oscillating star}\label{identifying}
The asteroseismic mass and radius are consistent with those from the ELC model for both stars. \newrevise{The surface gravity of the two stars from ELC are nearly identical, and both agree with the asteroseismic value. While neither star's mean density agrees with the asteroseismic value, Star 2 is slightly closer than Star 1. Since one of the scaling equations gives mean density independent of temperature and $\nu_{\rm max}$ (Equation \ref{density}),} one might na{\"i}vely expect a better asteroseismic estimation of density compared to surface gravity. \newrevise{It is therefore important to consider the temperature dependence of Equation \ref{gravity}.} From \citet{gau13}, \citet{gau14}, and the present work, asteroseismic masses and radii were reported to be $(1.7 \pm 0.3 \ M_\odot, 7.7 \pm 0.4 \ R_\odot)$, $(2.06 \pm 0.13 \ M_\odot, 8.10 \pm 0.18 \ R_\odot)$, and $(2.17 \pm 0.14 \ M_\odot, 8.26 \pm 0.18 \ R_\odot)$, respectively. Among these, $\nu_{\rm{max}}$ does not vary much ($102.2, 106.4, 106.4 \ \mu\rm{Hz}$), and $\Delta \nu$ varies even less ($8.3, 8.32, 8.31 \ \mu\rm{Hz}$), while the assumed temperatures were 4699 K (from the KIC), 4857 K (from \citealt{hub14.2}), and 5000 K (this work). Even if temperature is the least influential parameter in the asteroseismic scalings, we are at a level of precision where errors on temperature dominate the global asteroseismic results. \newrevise{In this case, while Star 2 appears to be a better candidate for the main oscillator at a glance, scaling relations alone cannot be used to prefer one star over the other. However, in Section \ref{actrot} we demonstrate that Star 2 is likely less active than Star 1. Based on this, we tentatively assign Star 2 as the main oscillator.}

\subsubsection{Surface gravity disagreement}\label{gravity_compare}
The asteroseismic $\log g$ measurement nearly agrees with those from ELC, yet all three are some 0.3 dex lower than the spectroscopic $\log g$ values, as can be seen in Table \ref{table2}. This discrepancy is similar to the difference found for giant stars by \citet{hol15}. They investigate a large sample of stars from the ASPCAP (APOGEE Stellar Parameters and Chemical Abundances Pipeline) which have $\log g$ measured via spectroscopy and asteroseismology. They find that spectroscopic surface gravity measurements are roughly 0.2--0.3 dex too high for core-He-burning (red clump) stars and roughly 0.1--0.2 dex too high for shell-H-burning (red giant branch) stars. \citet{hol15} speculate the difference may be partially due to a lack of treatment of stellar rotation, and derive an empirical calibration relation for a ``correct'' $\log g$ for red giant branch stars only. However, the stars in KIC 9246715 do not rotate particularly fast ($v_{\rm{rot}} \sin i \lesssim 8 \ \rm{km \ s}^{-1}$, which includes a contribution from macroturbulence as discussed in Section \ref{parameters}), so we cannot dismiss this discrepancy so readily.

\begin{figure}[ht!]
\begin{center}
\includegraphics[width=1\columnwidth]{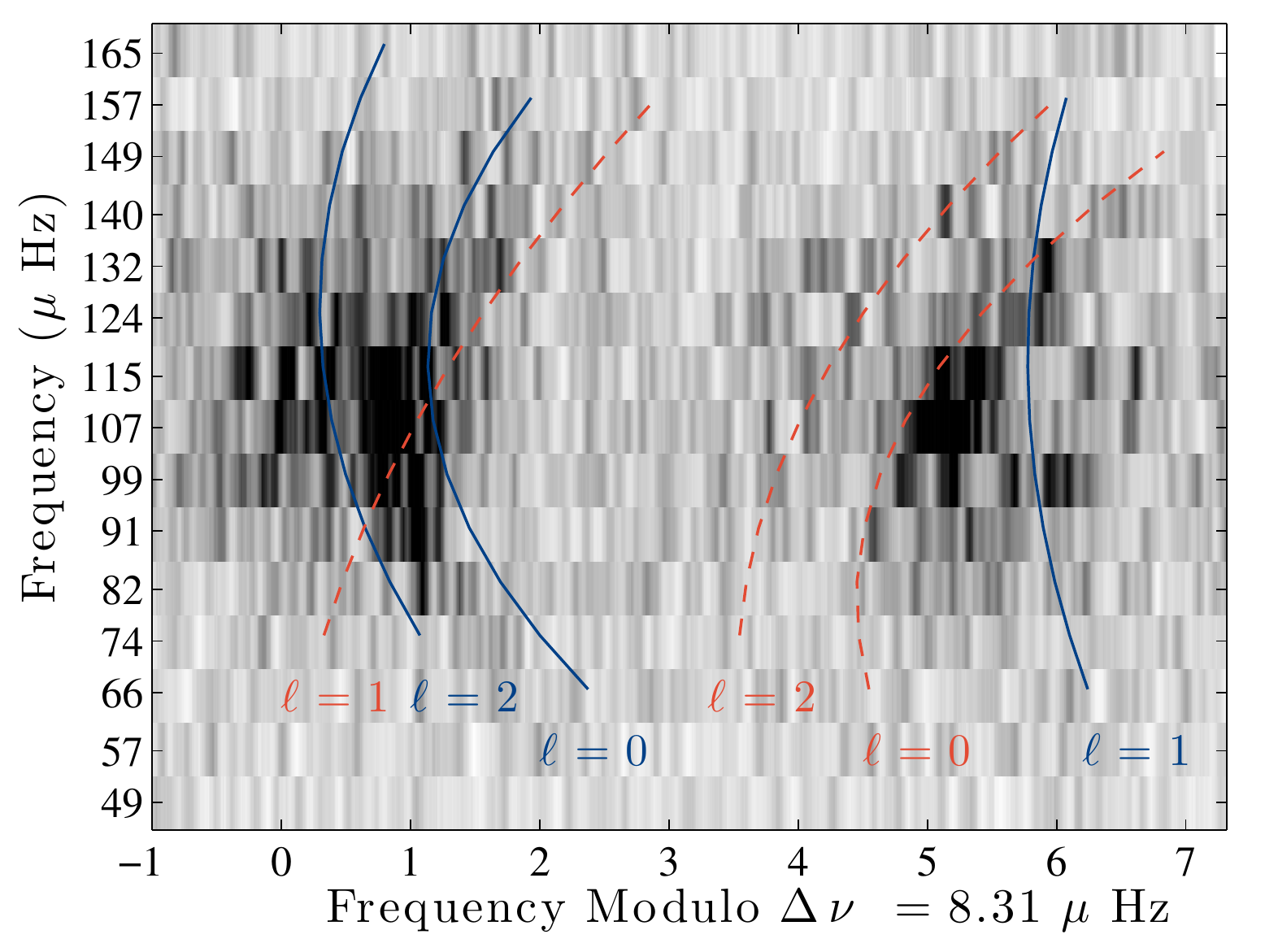}
\caption{\label{fig:echelle} \'Echelle diagram of KIC 9246715's power density spectrum. Darker regions correspond to larger peaks in power density. The power density spectrum is smoothed by a boxcar over seven bins and cut into 8.31-$\mu\rm{Hz}$ chunks; each is then stacked on top of its lower-frequency neighbor. This representation allows for visual identification of the modes. Lines are plotted to guide the eye toward a theoretical mode distribution according to the red giant universal pattern \citep{mos11}. It illustrates how we expect the modes to appear, but is not the result of a fit. Solid blue and dashed red lines are associated with the main (nominally Star 2) and marginally-detected (nominally Star 1) oscillations, respectively. The variable $\ell$ labels each mode by its spherical degree. Large spacing is $\Delta\nu = 8.31 \ \mu\rm{Hz}$ for the main (blue) lines and $8.60\ \mu\rm{Hz}$ for the marginal (red) lines (see Section \ref{subsec_second_osc}).
}
\end{center}
\end{figure}

\vspace{0.2em}

\subsection{A hint of a second set of oscillations}
\label{subsec_second_osc}
Given that the giants in KIC 9246715 are nearly twins, we test whether it is possible that we see only one set of oscillation modes because both stars are oscillating with virtually identical frequencies. \newrevise{The predicted $\nu_{\rm{max}}$ values for these} not-quite-identical stars are
$103.4\substack{+1.6 \\ -1.1}$ and $104.1\substack{+1.1 \\ -1.2} \ \mu\rm{Hz}$ 
for Star 1 and Star 2, respectively (from an inversion of Equations \ref{density} and \ref{gravity}), and the predicted $\Delta\nu_{\rm{obs}}$ are 
$8.14\substack{+0.06 \\ -0.03}$ and $8.20\substack{+0.03 \\ -0.04} \ \mu\rm{Hz}$ 
for Star 1 and Star 2, respectively.
\newrevise{As described in Section \ref{subsubsec_main_osc}, the intrinsic observed mode linewidths is $0.4 \ \mu \rm{Hz}$, which is about four times wider than expected. To quantify how likely it is for oscillation modes like this to overlap one another, we use the ELC model results from Section \ref{all-eclipse} to calculate distributions of expected $\Delta \nu$ for each star. We find that in 89\% of the cases, $|\Delta \nu_1 - \Delta \nu_2| < 0.4 \ \mu \rm{Hz}$. This suggests that, if both stars do indeed exhibit solar-like oscillations, some degree of mode overlap is likely.}

Searching for a second set of oscillations is motivated by the broad, mixed-mode-like appearance of the $\ell=0$ modes in Figure \ref{fig:echelle}, where mixed modes are not physically possible, and by the faint diagonal structure mostly present on the upper left side of the $\ell=1$ mode ridge. Even though oscillation modes from the two stars should not perfectly overlap, modes of degree $\ell=0,1$ of one star can almost overlap modes of degree $\ell=1,0$ of the other star.

The universal red giant oscillation pattern \citep{mos11} yields $\Delta \nu =  8.31 \pm 0.02 \ \mu\rm{Hz}$ for this system (Section \ref{subsubsec_main_osc}). \newrevise{However, it appears that the asymptotic relations for pressure-modes and mixed modes from the main oscillating star alone may not reproduce the position of all the peaks in the power spectrum.}
We therefore test the hypothesis of a binary companion. The universal oscillation pattern allows us to tentatively allocate the extra peaks to a pressure-mode oscillation pattern based on $\Delta \nu = 8.60 \pm 0.04 \ \mu\rm{Hz}$\footnote{The quoted uncertainty here is an ``internal'' error bar which assumes an underlying distribution of modes that corresponds to the red giant universal pattern.}. \newrevise{This putative oscillation spectrum is globally interlaced with the main oscillations,} with the dipole modes of one component close to the radial modes of the other component, and vice versa.

This value aligns the diagonal structure seen in the \'echelle diagram and satisfies the ($\ell=0,1-\ell=1,0$) near-overlap evident in Figure \ref{fig:echelle}. However, because these peaks are only marginally detected, $\nu_{\rm{max}}$ cannot be measured. The asteroseismic scaling connecting $\Delta\nu$ with the mean density yields $\bar{\rho}/\bar{\rho}_\odot = (4.14 \pm 0.02)\times 10^{-3}$. \revise{This density is larger than we expect; in fact, we expect Star 1 to be less dense than Star 2, the suspected main oscillator.} \newrevise{This casts further doubt on the second set of oscillations, and it may be a spurious detection.}

\newrevise{Finally,} we investigate whether the modes show any frequency modulation as a function of orbital phase by examining portions of the power spectrum spanning less than the orbital period. However, the solar-like oscillations modes are short-lived (about 23 days from an average $0.5 \ \mu\rm{Hz}$ width of $l=0$ modes), so it is difficult to clearly resolve Doppler-shifted modes in a power spectrum of a light curve segment. At $\nu_{\rm{max}} = 106 \ \mu \rm{Hz}$, the maximum frequency shift expected from a $60 \ \rm{km} \ \rm{s}^{-1}$ difference in radial velocity is $0.02 \ \mu \rm{Hz}$. This is less than the intrinsic mode line width, and therefore not observable.

\subsection{Signatures of stellar activity}\label{actrot}
KIC 9246715 is an interesting pair of well-separated red giants that exhibit photometric variations from stellar activity, weak or absent solar-like oscillations, and a notably eccentric orbit. In this and the following section, we discuss how stellar activity and tidal forces have acted over the binary's lifetime to arrive at the system we see today. \revise{The first confirmed case of activity and/or tides suppressing convection-driven oscillations was \citet{der11}, and as \citet{gau14} showed, stellar activity and tides likely play an important role in many RG/EBs.}

\revise{In this system, the light curve residuals discussed in Section \ref{segment} and Figure \ref{fig:ELCresult} show significant scatter during both eclipses, and especially primary eclipse (when Star 1 is in front).} This means at least Star 1 is magnetically active, and activity in the system is further supported by photometric variability of up to 2\% on a timescale approximately equal to half the orbital period \citep{gau14}. A magnetically active Star 1 is also consistent with Star 2 as the suspected main oscillator, because strong magnetic fields may be responsible for damping solar-like oscillations, \newrevise{as described in \citet{ful15}}.

\begin{figure*}[h!]
\begin{center}
\includegraphics[width=2.3\columnwidth]{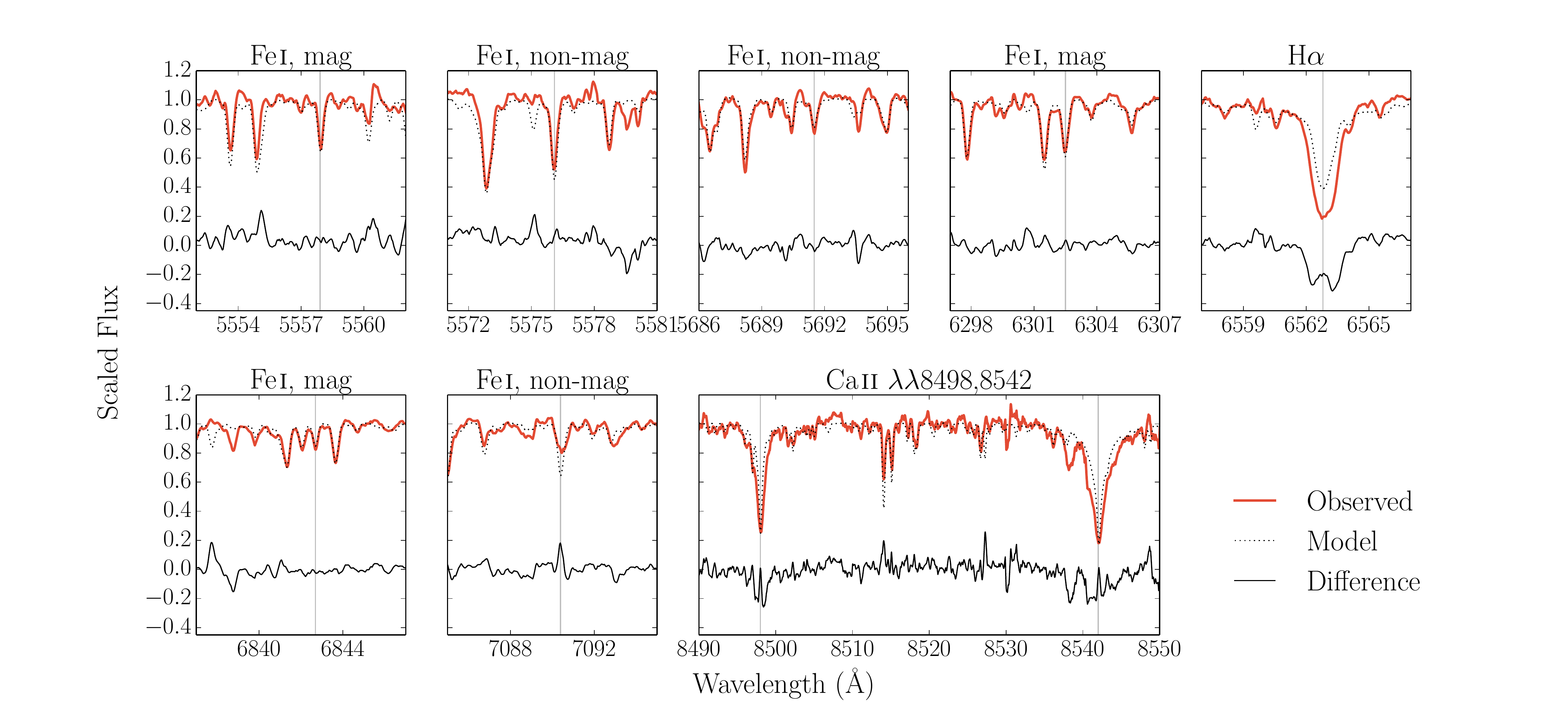}
\caption{\label{fig:emission1} \emph{Top of each panel:} observed FDBinary-extracted spectrum of Star 1 (red) together with a stellar template (dotted black line). \emph{Bottom of each panel:} difference between the observed and model spectra. Vertical lines show the position of each absorption line. Broadened magnetic-sensitive lines would indicate Zeeman splitting, but this is not observed. Net emission in the $\rm{H}\alpha$ and {\rm Ca}\kern 0.1em{\sc ii} lines is a characteristic signature of chromospheric magnetic activity, but this is not observed either. Instead, the $\rm{H}\alpha$ line is deeper and broader than the model.
}
\end{center}
\end{figure*}

\begin{figure*}[h!]
\begin{center}
\includegraphics[width=2.3\columnwidth]{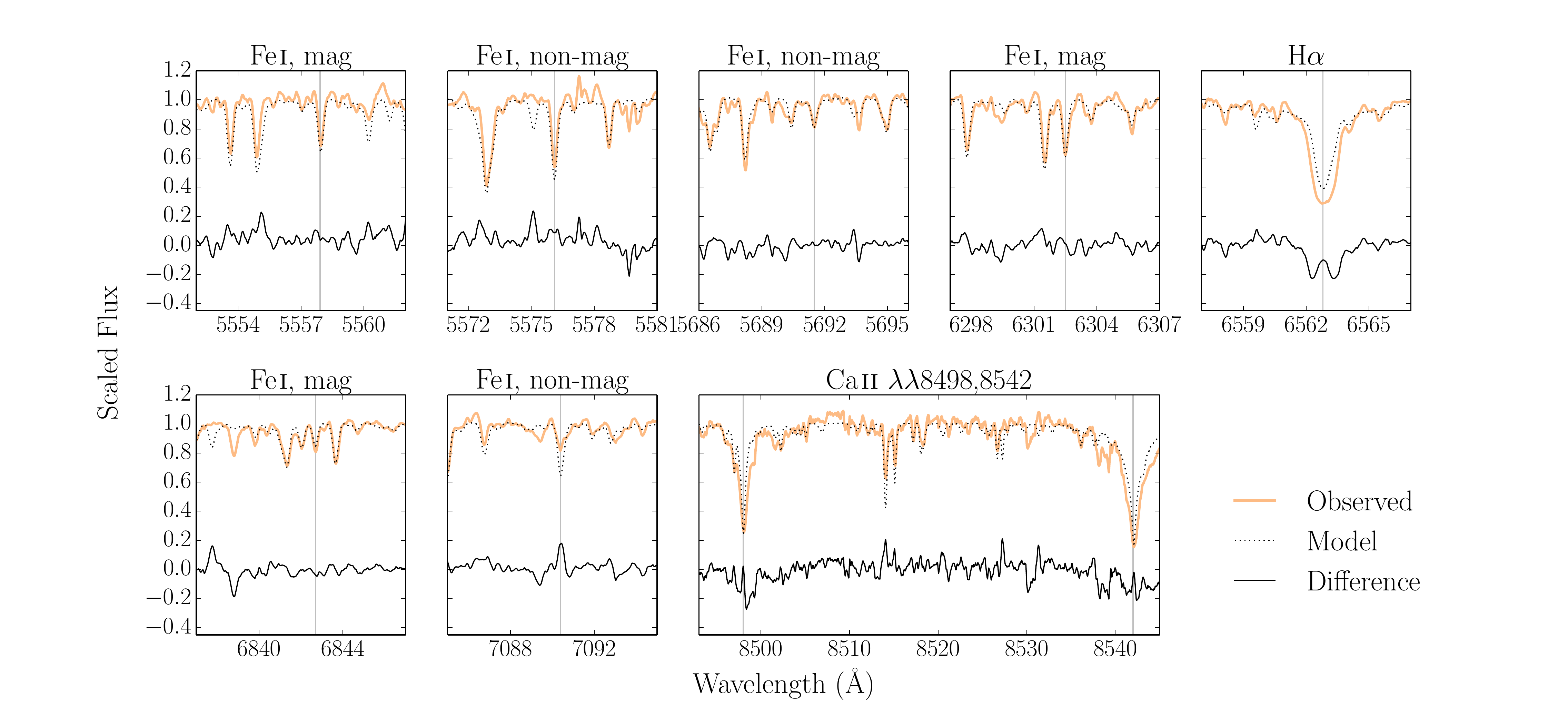}
\caption{\label{fig:emission2} The same as in Figure \ref{fig:emission1}, but for Star 2 (yellow). No signatures of Zeeman broadening or chromospheric emission are present. The $\rm{H}\alpha$ absorption is slightly deeper and broader than expected, but not as much as that of Star 1.
}
\end{center}
\end{figure*}

Figures \ref{fig:emission1} and \ref{fig:emission2} investigate whether magnetic activity has any appreciable effect on absorption lines in either star. Following the approach of \citet{fro12}, we plot each target spectrum (solid colored line) on top of a model (dotted line), and show the difference below (solid black line). The model spectrum is a PHOENIX BT-Settl stellar atmosphere like the one described in Section \ref{bf} \citep{all03,asp09}, with $T_{\rm{eff}} = 5000$ and $\log g = 3.0$. It has been convolved to a lower resolution much closer to that of the ARCES and TRES spectrographs.

We examine a selection of the strongest {\rm Fe}\kern 0.1em{\sc i} lines which fall in the disentangled wavelength region and are either prone to Zeeman splitting in the presence of strong magnetic fields \citep{har73}, or not \citep{sis70}. The non-magnetic lines serve as a control. We find none of the six panels of {\rm Fe}\kern 0.1em{\sc i} absorption lines in either star show any significant deviation from the model spectrum. Thus, there is no apparent Zeeman broadening, which is unsurprising for evolved red giants. Magnetic fields must be quite strong to produce this effect. However, the $\rm{H}\alpha$ and {\rm Ca}\kern 0.1em{\sc ii} absorption lines, which can be indicators of chromospheric activity, are somewhat more interesting. The $\rm{H}\alpha$ line appears significantly deeper and broader than the model in both stars. While net emission is typically associated with activity, \citet{rob90} show several examples of main sequence stars with increased $\rm{H}\alpha$ absorption due to chromospheric heating, although they caution it is difficult to separate the photospheric and chromospheric contributions to the line. Still, the increased $\rm{H}\alpha$ absorption equivalent width is slightly more pronounced in Star 1 than Star 2. \revise{While this may not be a significant difference on its own, taken together \newrevise{with the increased scatter in the primary eclipse residuals from Figure \ref{fig:ELCresult}}, it also suggests Star 1 is the more magnetically active of the pair.} It is unclear whether the {\rm Ca}\kern 0.1em{\sc ii} doublet shows signs of excess broadening or increased equivalent width, but these lines certainly do not have \emph{smaller} equivalent widths than the model.

\revise{The overall photometric variability from Figure \ref{fig:lcfig2} and increased scatter in the primary eclipse residual from Figure \ref{fig:ELCresult} indicate that both stars are moderately magnetically active, and Star 1 more so than Star 2. This is consistent with increased $\rm{H}\alpha$ absorption in both stars (and especially Star 1),} and supports our suspicion that Star 2 is the main oscillator, and that stellar activity is suppressing solar-like oscillations in Star 1.

\subsection{Stellar evolution and tidal forces}\label{tides}
\begin{figure*}[ht!]
\begin{center}
\includegraphics[width=1.95\columnwidth]{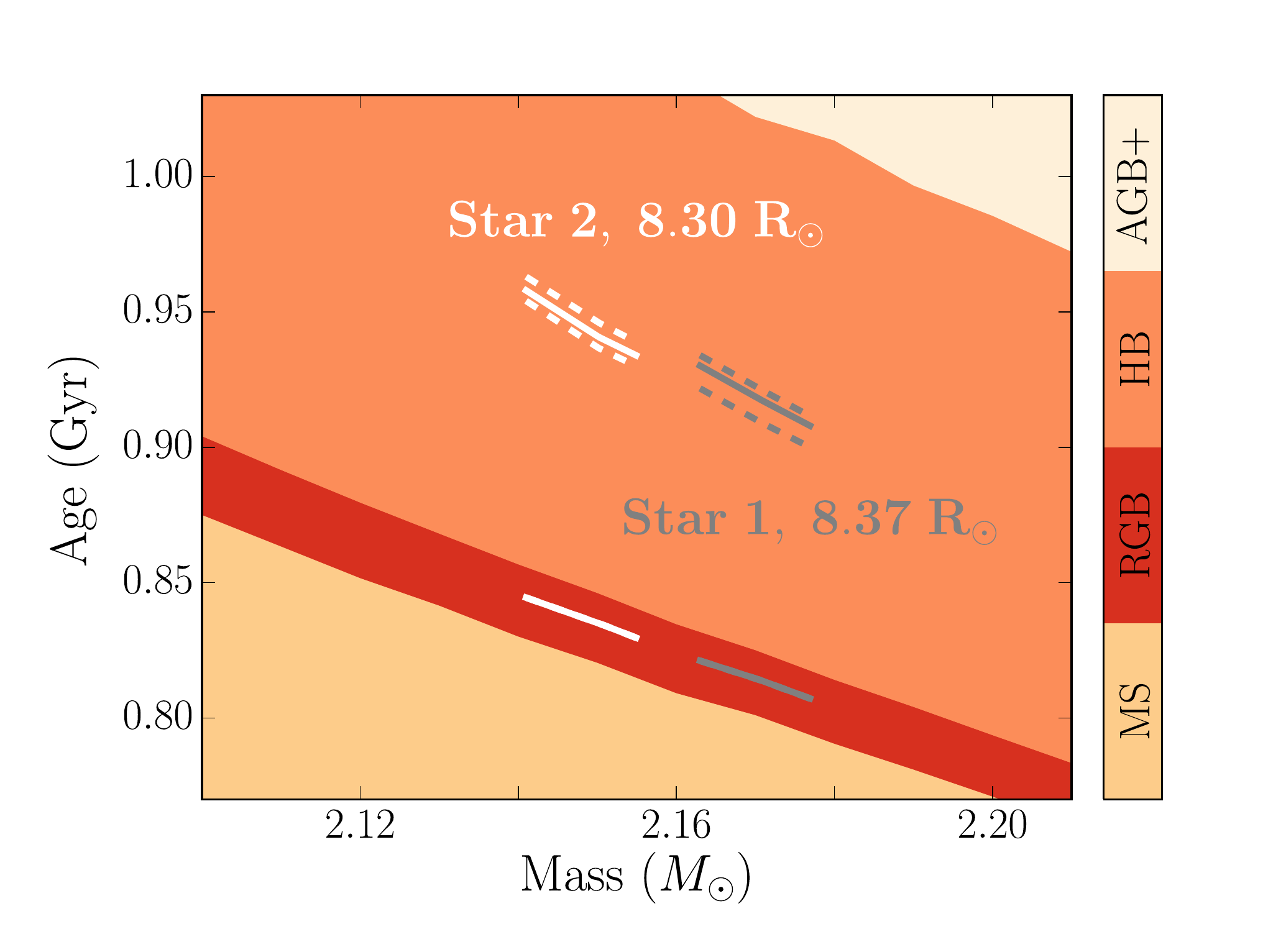}
\caption{\label{fig:mesa} Ages for a suite of MESA stellar evolution models for stars of different masses. Color indicates the evolutionary state of a star as it moves from the Main Sequence (MS) $\rightarrow$ Red Giant Branch (RGB) $\rightarrow$ Secondary Red Clump/Horizontal Branch (HB) $\rightarrow$ Asymptotic Giant Branch and beyond (AGB+). Lines of constant radius equal to $R_1$ and $R_2$ that fall within the one-sigma errors in mass are shown (gray, $R_1 \pm \sigma_{R_1}$ corresponding to $M_1 \pm \sigma_{M_1}$; white, $R_2 \pm \sigma_{R_2}$ corresponding to $M_2 \pm \sigma_{M_2}$). All models assume a mixing-length parameter of $\alpha = 2.5$. It is possible for both stars in KIC 9246715 to be the same age on the HB, but not on the RGB.
}
\end{center}
\end{figure*}

Over the course of KIC 9246715's life, both stars have evolved in tandem to reach the configuration we see today. We quantify this with simple stellar evolution models created using the Modules for Experiments in Stellar Astrophysics (MESA) code \citep{pax11,pax13,pax15}. Figure \ref{fig:mesa} presents a suite of models with various initial stellar masses. All the models include overshooting for all the convective zone boundaries with an efficiency of $f = 0.016$ \citep{her00}, assume no mass loss, \revise{and set the mixing-length parameter $\alpha = 2.5$. The standard solar value of $\alpha = 2$ does not allow for sufficiently small stars beyond the red giant branch}. The stage of each model star's life as it ages in Figure \ref{fig:mesa} is color-coded, and \revise{curved lines of constant radii corresponding to $R_1 \pm \sigma_{R_1}$ (gray) and $R_2 \pm \sigma_{R_2}$ (white), within the ranges of $M_1 \pm \sigma_{M_1}$ and $M_2 \pm \sigma_{M_2}$, respectively, are shown. There are two instances in each pair of model stars' lives when they have the same radii as the stars in KIC 9246715: once on the red giant branch, and again on the secondary red clump (horizontal branch).}

In general, coeval stars on the red giant branch must have masses within $1\%$ of each other, \revise{whereas masses can differ more on the horizontal branch due to its longer evolutionary lifetime. Both model stars in Figure \ref{fig:mesa} can be the same age on the horizontal branch, but not on the red giant branch.
\newrevise{Stars 1 and 2 in Figure \ref{fig:mesa} have red giant branch ages of $8.13\substack{+0.08 \\ -0.06} \times 10^8 \ \rm{yr}$ and $8.36\substack{+0.08 \\ -0.06} \times 10^8 \ \rm{yr}$, respectively, and horizontal branch ages of $9.17 \pm 0.17 \times 10^8 \ \rm{yr}$ and $9.42\substack{+0.20 \\ -0.13} \times 10^8 \ \rm{yr}$, respectively.}
Without $\alpha > 2$, the MESA model stars on the horizontal branch are always larger than those in KIC 9246715. We consider several ideas as to why the MESA models and the evolutionary stage determined from asteroseismic mixed-mode period spacing in Section \ref{subsubsec_mixed} may differ:}
\begin{itemize}
\item Mass loss: Adding a prescription for red-giant-branch mass loss ($\eta = 0.4$, \revise{a commonly adopted value of the parameter describing mass-loss efficiency}, see \citealt{mig12}) to the MESA model does not appreciably change stellar radius as a function of evolutionary stage. Even a more extreme mass-loss rate ($\eta = 0.7$) does not significantly affect the radii, essentially because the star is too low-mass to lose much mass.
\item He abundance: Increasing the initial He fraction in the MESA model does not allow for smaller stars in the red clump phase, because shell-H burning is very efficient with additional He present. As a result, the star maintains a high luminosity and therefore a larger radius as it evolves from the tip of the red giant branch to the red clump.
\item Convective overshoot: The MESA models in this work assume a reasonable overshoot efficiency as described above ($f = 0.016$). We tried varying this from 0--0.03, and can barely make a red clump star as small as $8.3 \ R_\odot$ when $f = 0.01$. With less overshoot, the RGB phase as shown in Figure \ref{fig:mesa} increases in duration, which allows a higher probability for stars of $M_1$ and $M_2$ to both be on the RGB.
\item Period spacing: The period spacing $\Delta \Pi_1 = 150 \ \rm{s}$ may not be measuring what we expect due to rotational splitting of mixed oscillation modes. If the true period spacing is closer to $\Delta \Pi_1 \simeq 80 \ \rm{s}$, \revise{this would put the oscillating star on the red giant branch. However, as demonstrated in Section \ref{subsubsec_mixed}, the mixed modes do agree best with a secondary red clump star.} A detailed discussion of rotational splitting behavior in slowly rotating red giants is explored in \citet{gou13}.
\item Mixing length: \revise{As discussed above,} increasing the mixing-length parameter from the standard solar value of $\alpha = 2$ to $\alpha = 2.5$ in the MESA model, which effectively increases the efficiency of convection, produces a red clump star small enough to agree with both measured radii. This is because it reduces the temperature gradient in the near-surface layers, increasing the effective temperature while reducing the radius at constant luminosity. \revise{This is what we employ to make horizontal branch stars that agree with $R_1$ and $R_2$.}
\end{itemize}

Beyond a stellar evolution model, it is important to consider how each star has affected the other over time. When the two stars in KIC 9246715 reach the tip of the red giant branch, they have radii of approximately $25 \ R_\odot$, which is still significantly smaller than the periastron separation ($r_{\rm{peri}} = (1-e)~a = 137 \ R_\odot$). We never expect the stars to experience a common envelope phase, so this cannot be used to constrain the present evolutionary state.

To estimate how tidal forces change orbital eccentricity, we follow the approach of \citet{ver95}. They use a theory of the equilibrium tide first proposed by \citet{zah77} to calculate a timescale for orbit circularization as a star evolves. It is important to note that \citet{ver95} assumed circularization would proceed by a small secondary star (main sequence or white dwarf) imposing an equilibrium tide on a large giant, while the situation with KIC 9246715 is more complicated. For a thorough review of tidal forces in stars, see \citet{ogi14}.

From Equation 2 in \citet{ver95}, the timescale $\tau_c$ on which orbital circularization occurs is given by

\begin{eqnarray}
\frac{1}{\tau_c} & \equiv &
\frac{{\rm{d}} \ln e} {{\rm{d}} t} \nonumber \\
& \simeq & -1.7 {\left( \frac{T_{\rm{eff}}}{4500 \rm{K}} \right)}^{4/3} \left( \frac{M_{\rm{env}}}{M_{\odot}} \right)^{2/3} \\
& \times & \ \frac{M_{\odot}}{M} \frac{M_2}{M} \frac{M+M_2}{M} \left( \frac{R}{a} \right)^8 \ \rm{yr}^{-1}, \nonumber
\end{eqnarray}

\noindent where $M$, $R$, and $T_{\rm{eff}}$ are the mass, radius, and temperature of a giant star with dissipative tides, $M_{\rm{env}}$ is the mass of its convective envelope, $M_2$ is the mass of the companion star, and $a$ is the semi-major axis of the binary orbit.

We integrate this expression over the lifetime of KIC 9246715 to estimate the total expected change in orbital eccentricity, $\Delta \ln e$. We assume $a$ is constant and that there is no mass loss. Because KIC 9246715 is a detached binary, we can separate the integral into a part that is independent of the orbit and a part that must be integrated over time:

\begin{eqnarray}\label{tide1}
\Delta \ln e  & = &
\int_0^t \frac{\rm{d}t'}{\tau_c(t')} \nonumber \\
& \simeq & -1.7 \times 10^{-5} {\left( \frac{M}{M_{\odot}} \right)}^{-11/3} \\
& \times & \ q(1+q)^{-5/3} \ I(t) {\left( \frac{P_{\rm{orb}}}{\rm{day}} \right)}^{-16/3}, \nonumber
\end{eqnarray}

\noindent where $q$ is the mass ratio and

\begin{equation}
I(t) \equiv \int_0^t \left( \frac{T_{\rm{eff}}(t')}{4500 \rm{K}} \right)^{4/3} \left( \frac{M_{\rm{env}}(t')}{M_{\odot}} \right)^{2/3} \left( \frac{R(t')}{R_{\odot}} \right)^8 dt'. \nonumber
\end{equation}

For the MESA model described above with $M = 2.15 \ M_{\odot}$, \revise{we compute $\Delta \ln e = -2.3 \times 10^{-5}$ up until $t = 8.3 \times 10^8$ and $\Delta \ln e = -0.17$ up until $t = 9.4 \times 10^8$ years (the ages corresponding to $R \simeq 8.3 \ R_{\odot}$). Rewriting these as $\log [-\Delta \ln e] = -4.6$ and $\log [-\Delta \ln e] = -0.77$,} which are both less than zero, indicates that the binary has \emph{not} had sufficient time to circularize its orbit, though it is possible the system's initial eccentricity was higher than the $e = 0.35$ we observe today.

The two stars in KIC 9246715 have very similar masses, radii, and temperatures, so this rough calculation is valid both for Star 1 acting on Star 2 and vice versa. Given more time to evolve past the tip of the red giant branch and well onto the red clump (with $R \simeq 25 \ R_\odot$ for the second time), $\log [-\Delta \ln e]$ becomes greater than zero and the expectation is a circular orbit. Therefore, the observed eccentricity is consistent with \revise{both a red giant branch star aged approximately $8.3 \times 10^8$ years and with a secondary red clump star just past the tip of the red giant branch aged approximately $9.4 \times 10^8$ years.}

Tidal forces also tend to synchronize a binary star's orbit with the stellar rotation period, generally on shorter timescales than required for circularization \citep{ogi14}. Hints of KIC 9246715's stellar rotation behavior are present throughout this study: quasi-periodic light curve variability on the order of half the orbital period, \revise{residual scatter between light curve observations and the best-fit model during both eclipses}, a constraint on $v_{\rm{rot}} \sin i$ from spectra, and an asteroseismic period spacing consistent with a red clump star yet not clear enough to measure a robust core rotation rate.

While full tidal circularization has not occurred, it is clear that modest tidal forces have played a role in the evolution of KIC 9246715, and may be linked to the absence or weakness of solar-like oscillations. Future studies of RG/EBs with different evolutionary histories and orbital configurations will help explore this connection further.

\section{Conclusions}\label{conclude}

We have characterized the double red giant eclipsing binary KIC 9246715 with a combination of dynamical modeling, stellar atmosphere modeling, and global asteroseismology, and have investigated the roles of magnetic activity, tidal forces, and stellar evolution in creating the system we observe today. KIC 9246715 represents a likely future state of similar-mass RG/EB systems and raises interesting questions about the interactions among stellar activity, tides, and solar-like oscillations.

The two stars in KIC 9246715 are nearly twins ($M_1 = 2.171\substack{+0.006 \\ -0.008} \ M_{\odot}$, $M_2 = 2.149\substack{+0.006 \\ -0.008} \ M_{\odot}$, $R_1 = 8.37\substack{+0.03 \\ -0.07} \ R_{\odot}$, $R_2 = 8.30\substack{+0.04 \\ -0.03} \ R_{\odot}$), yet we find only one set of solar-like oscillations strong enough to measure robustly ($M = 2.17 \pm 0.14 \ M_{\odot}$, $R = 8.26 \pm 0.18 \ R_{\odot}$). \newrevise{The asteroseismic mass and radius agree with both Star 1 and Star 2, as does the surface gravity derived from asteroseismology ($\log g = 2.942 \pm 0.008$; compare with $\log g_1 = 2.929\substack{+0.007 \\ -0.003}$ and $\log g_2 = 2.932\substack{+0.003 \\ -0.004}$). The asteroseismic density, which is not a function of effective temperature, is systematically larger than Star 1 and Star 2, but is a slightly closer match with Star 2 ($\bar{\rho}/\bar{\rho}_\odot = (3.86 \pm 0.02) \times 10^{-3}$; compare with $\bar{\rho}_1/\bar{\rho}_\odot = (3.70\substack{+0.04 \\ -0.09}) \times 10^{-3}$ and $\bar{\rho}_2/\bar{\rho}_\odot = (3.76\substack{+0.06 \\ -0.04}) \times 10^{-3}$). As a result, we cannot conclude which star is the source of the main oscillations from asteroseismology alone. However, Star 2 appears to be less active than Star 1, and we therefore tentatively assign the main oscillations to Star 2. The modes are four times wider than expected with amplitudes only $\sim 60\%$ as high as those in red giants with similar global oscillation properties, likely due to a combination of overlapping adjacent modes and magnetic damping.} We identify a second set of marginally detectable oscillations potentially attributable to Star 1, for which only $\Delta \nu$ can be estimated, yielding a higher average density than the main oscillation spectrum. \newrevise{This is not consistent with the expected density of Star 1, however, which is less than that of Star 2. These extra modes may represent a spurious detection.}

Surface gravities from dynamical modeling and asteroseismology nearly agree, while surface gravities from stellar atmosphere modeling are higher ($\log g_1 = 3.21 \pm 0.45$, $\log g_2 = 3.33 \pm 0.37$). A similar discrepancy has been found between the asteroseismic and spectroscopic surface gravities of other giant stars, but the physical cause is unknown. \revise{Radii from stellar evolution models are consistent with a pair of nearly-coeval stars either on the red giant branch with an age of approximately $8.3 \times 10^8$ years, or coeval stars on the horizontal branch with an age of about $9.4 \times 10^8$ years. However, the period spacing of mixed oscillation modes clearly indicates that the main oscillator in KIC 9246715 is on the secondary red clump, and we conclude that KIC 9246715 is a pair of secondary red clump stars.}

Red giants are ideal tools for probing the Milky Way Galaxy via asteroseismology, so it is crucial that we understand the accuracy and precision of asteroseismically-derived physical parameters. Along the same lines, more than half of cool stars should be in binary or multiple systems, so galactic studies must be done carefully due to external influences of binarity on solar-like oscillations. Detailed studies of the handful of known RG/EBs are crucial to ensure we understand these galactic beacons. Future work will characterize the other known oscillating RG/EBs as well as several non-oscillating RG/EBs. These have the potential to become some of the best-studied stars while simultaneously helping us better understand the structure of the Milky Way.

\vspace{3em}

\acknowledgments
This paper was written collaboratively on the web with Authorea at \url{authorea.com/2409}. M. L. R. thanks the New Mexico Space Grant Council for support, D. Chojnowski for assistance with APOGEE, D. Muna for the SciCoder workshop, and L. C. Mayorga for programming assistance. J. J. acknowledges support from NASA ADAP grant NNX14AR85G. E. C. \& P. B. received funding from the European Community's Seventh Framework Programme [FP7/2007--2013] under grant agreements 312844 (SPACEINN) and 269194 (IRSES/ASK). P. B. also received funding from CNES grants at CEA. D. W. L. acknowledges partial support from NASA's \emph{Kepler} Mission under Cooperative Agreement NNX13AB58A with the Smithsonian Astrophysical Observatory. We thank Leo Girardi and the anonymous referee for valuable feedback and Kresimir Pavlovski for useful discussions. This paper uses data from the Apache Point Observatory 3.5-meter telescope, which is owned and operated by the Astrophysical Research Consortium, and data collected by the \emph{Kepler} mission, which is funded by the NASA Science Mission directorate. This research made use of Astropy \citep{astropy}, PyAstronomy (\url{github.com/sczesla/PyAstronomy}), PyKE \citep{pyke}, NASA's ADS Bibliographic Services, and the AstroBetter blog and wiki.
\\

{\it Facilities:} \facility{Kepler}, \facility{APO:3.5m (ARCES)}, \facility{FLWO:1.5m (TRES)}, \facility{Sloan (APOGEE)}

\vspace{3em}

\appendix
\section{Oscillation modes fit with DIAMONDS}
\label{appendix}

\begin{figure*}[ht!]
\begin{center}
\includegraphics[width=2\columnwidth]{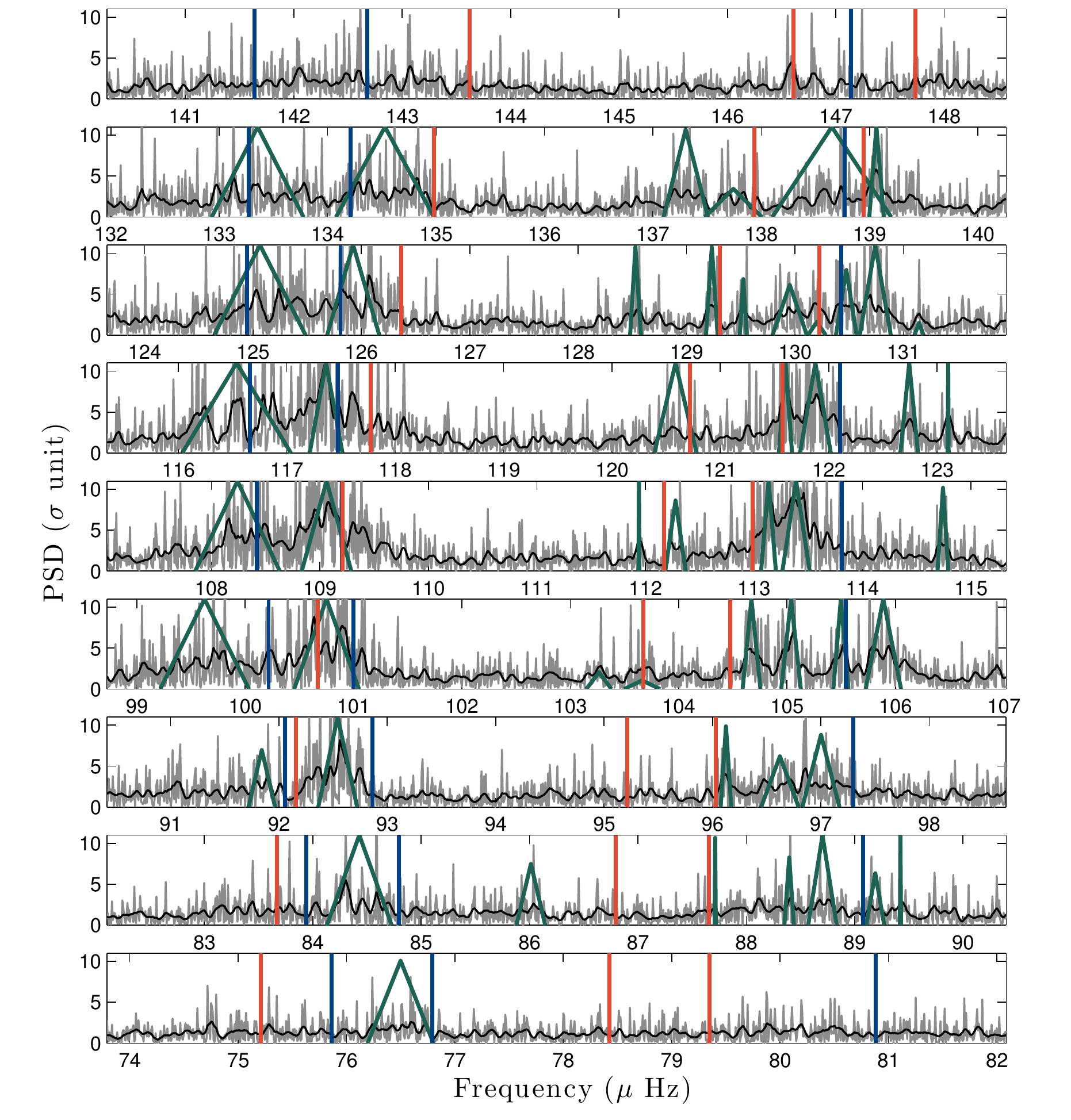}
\caption{\label{fig:appendixfig} Power spectral density (PSD) of KIC 9246715 as function of frequency, in the form of an \'echelle diagram (compare to Figure \ref{fig:echelle}). The PSD has been whitened, or divided by the background fitting, which casts the y-axis in terms of sigma. Solid blue lines indicate the universal pattern for an oscillator with $\Delta \nu = 8.31 \ \mu \rm{Hz}$ (main oscillator), while red lines indicate the same for $\Delta \nu = 8.60 \ \mu \rm{Hz}$ (marginal detection). Dark green triangles correspond to the location of fit peaks from \textsc{D\large{iamonds}} (Table \ref{appendixtable}). The width of each triangle's base is the mode linewidth from the fit, and taller triangles represent higher detection confidence.
}
\end{center}
\end{figure*}

In this appendix, we present the frequencies fit by \textsc{D\large{iamonds}} \citep{cor14}, as described in Section \ref{subsubsec_mixed}. We follow the methodology for the peak bagging analysis of a red giant star in \citet{cor15}. Each fit mode's frequency together with its angular degree $\ell$, azimuthal order $m$, amplitude or height, linewidth (when applicable), and probability of detection is listed in Table \ref{appendixtable}. Figure \ref{fig:appendixfig} shows these modes superimposed on the power density spectrum of KIC 9246715, which is split up like an \'echelle diagram for clarity. For comparison, we also plot the locations of where modes should fall according to the asymptotic relation \citep{mos12} for the main set of oscillations ($\Delta \nu = 8.31 \ \mu \rm{Hz}$) and the marginally detected second set of oscillations ($\Delta \nu = 8.60 \ \mu \rm{Hz}$). The power spectrum is quite noisy overall, exhibits wide modes with low amplitudes, and is challenging to interpret unambiguously. For a full discussion, see Section \ref{discuss}.

\begin{deluxetable*}{lcccc}
\tablecolumns{5}
\tablewidth{0pt}
\tabletypesize{\footnotesize}
\tablecaption{Oscillation modes in KIC 9246715 fit with DIAMONDS.}
\centering
\tablehead{
\colhead{Frequency} & \colhead{$(\ell, m)$} & \colhead{Amplitude or Height\tablenotemark{a}} & \colhead{Linewidth} & \colhead{Detection Probability\tablenotemark{b}} \\
\colhead{($\mu \rm{Hz}$)} & \colhead{} & \colhead{(ppm) or (ppm$^2 \ \mu \rm{Hz}^{-1}$)} & \colhead{($\mu \rm{Hz}$)} & \colhead{}
}
\startdata
 $76.50  \pm 0.01$  &  (0,  0)   &    $5.2 \pm 0.4$  &    $0.61  \pm 0.05$   &     0.91 \\
 $84.43  \pm 0.02$  &  (0,  0)   &   $10.9 \pm 0.4$  &    $0.60  \pm 0.05$   &     1.00 \\
 $92.54  \pm 0.01$  &  (0,  0)   &   $13.2 \pm 0.3$  &    $0.36  \pm 0.02$   &     1.00 \\
$100.75  \pm 0.02$  &  (0,  0)   &   $15.7 \pm 0.6$  &    $0.60  \pm 0.07$   &     1.00 \\
$109.06  \pm 0.01$  &  (0,  0)   &   $14.6 \pm 0.4$  &    $0.46  \pm 0.03$   &     1.00 \\
$117.37  \pm 0.01$  &  (0,  0)   &   $12.6 \pm 0.4$  &    $0.30  \pm 0.02$   &     1.00 \\
$125.92  \pm 0.04$  &  (0,  0)   &    $9.8 \pm 0.7$  &    $0.48  \pm 0.07$   &     1.00 \\
$134.53  \pm 0.02$  &  (0,  0)   &    $9.1 \pm 0.7$  &    $0.91  \pm 0.07$   &     1.00 \\
 $87.714 \pm 0.001$ &  (1,  ?)   &  $402\substack{+11 \\ -12}$  &  \nodata   &     0.97 \\
 $88.40  \pm 0.01$  &  (1, -1)   &    $1.6 \pm 0.1$  &    $0.088 \pm 0.006$  &     0.75 \\
 $88.70  \pm 0.01$  &  (1,  1)   &    $5.0 \pm 0.2$  &    $0.26  \pm 0.02$   &     0.99 \\
 $89.19  \pm 0.01$  &  (1, -1)   &    $1.7 \pm 0.1$  &    $0.17  \pm 0.01$   &     0.585 \\
 $89.422 \pm 0.001$ &  (1,  1)   &  $461\substack{+11 \\ -12}$  &  \nodata   &     0.99 \\
 $96.12  \pm 0.01$  &  (1,  1)   &    $2.8 \pm 0.5$  &    $0.10  \pm 0.01$   &     0.90 \\
 $96.62  \pm 0.02$  &  (1, -1)   &    $7.6 \pm 1.0$  &    $0.35  \pm 0.06$   &     0.56 \\
 $97.00  \pm 0.03$  &  (1,  1)   &    $7.9 \pm 1.0$  &    $0.34  \pm 0.05$   &     0.80 \\
$103.26  \pm 0.01$  &  (1, -1)   &    $3.7 \pm 0.2$  &    $0.23  \pm 0.02$   &     0.19 \\
$103.66  \pm 0.01$  &  (1,  1)   &    $6.3 \pm 0.4$  &    $0.33  \pm 0.03$   &     0.10 \\
$104.67  \pm 0.01$  &  (1, -1)   &    $6.1 \pm 0.3$  &    $0.17  \pm 0.01$   &     1.00 \\
$105.04  \pm 0.01$  &  (1,  1)   &    $8.7 \pm 0.4$  &    $0.18  \pm 0.02$   &     1.00 \\
$105.50  \pm 0.01$  &  (1, -1)   &    $5.9 \pm 0.3$  &    $0.14  \pm 0.01$   &     0.99 \\
$105.89  \pm 0.01$  &  (1,  1)   &    $8.4 \pm 0.5$  &    $0.33  \pm 0.03$   &     1.00 \\
$111.940 \pm 0.001$ &  (1, -1)   &  $435\substack{+16 \\ -33}$  &  \nodata   &     0.99 \\
$112.28  \pm 0.01$  &  (1,  1)   &    $3.5 \pm 0.3$  &    $0.19  \pm 0.02$   &     0.79 \\
$113.13  \pm 0.01$  &  (1, -1)   &    $7.7 \pm 0.4$  &    $0.14  \pm 0.01$   &     1.00 \\
$113.39  \pm 0.01$  &  (1,  1)   &   $12.3 \pm 0.5$  &    $0.25  \pm 0.02$   &     1.00 \\
$114.74  \pm 0.01$  &  (1,  ?)   &    $2.9 \pm 0.2$  &    $0.01  \pm 0.01$   &     0.93 \\
$120.59  \pm 0.03$  &  (1,  1)   &    $5.4 \pm 0.7$  &    $0.39  \pm 0.10$   &     0.99 \\
$121.60  \pm 0.01$  &  (1, -1)   &    $6.8 \pm 0.6$  &    $0.12  \pm 0.02$   &     0.99 \\
$121.88  \pm 0.02$  &  (1,  1)   &    $9.6 \pm 0.6$  &    $0.28  \pm 0.04$   &     1.00 \\
$122.74  \pm 0.02$  &  (1, -1)   &    $4.3 \pm 0.4$  &    $0.16  \pm 0.03$   &     0.99 \\
$123.101 \pm 0.003$ &  (1,  1)   &  $347\substack{+36 \\ -29}$  &  \nodata   &     1.00 \\
$128.53  \pm 0.01$  &  (1,  ?)   &    $3.2 \pm 0.3$  &    $0.10  \pm 0.01$   &     0.98 \\
$129.23  \pm 0.01$  &  (1, -1)   &    $3.7 \pm 0.4$  &    $0.11  \pm 0.01$   &     0.98 \\
$129.52  \pm 0.02$  &  (1,  1)   &    $1.3 \pm 0.1$  &    $0.07  \pm 0.01$   &     0.62 \\
$129.95  \pm 0.02$  &  (1, -1)   &    $6.0 \pm 0.3$  &    $0.32  \pm 0.05$   &     0.56 \\
$130.20  \pm 0.01$  &  (1,  1)   &    $4.8 \pm 0.3$  &    $0.16  \pm 0.02$   &     0.15 \\
$130.47  \pm 0.02$  &  (1, -1)   &    $3.9 \pm 0.3$  &    $0.19  \pm 0.03$   &     0.72 \\
$130.74  \pm 0.02$  &  (1,  1)   &    $6.1 \pm 0.5$  &    $0.29  \pm 0.05$   &     0.99 \\
$131.14  \pm 0.02$  &  (1, -1)   &    $1.7 \pm 0.1$  &    $0.08  \pm 0.01$   &     0.13 \\
$137.30  \pm 0.03$  &  (1, -1)   &    $5.0 \pm 0.7$  &    $0.41  \pm 0.13$   &     0.97 \\
$137.74  \pm 0.07$  &  (1,  1)   &    $3.3 \pm 0.8$  &    $0.53  \pm 0.18$   &     0.31 \\
$138.65  \pm 0.04$  &  (1, -1)   &    $7.7 \pm 0.9$  &    $1.10  \pm 0.26$   &     1.00 \\
$139.06  \pm 0.02$  &  (1,  1)   &    $4.2 \pm 0.5$  &    $0.13  \pm 0.03$   &     0.99 \\
 $91.84  \pm 0.01$  &  (2,  0)   &    $1.7 \pm 0.1$  &    $0.25  \pm 0.02$   &     0.63 \\
 $99.63  \pm 0.04$  &  (2,  0)   &   $11.1 \pm 1.0$  &    $0.82  \pm 0.11$   &     1.00 \\
$108.24  \pm 0.02$  &  (2,  0)   &   $11.6 \pm 1.2$  &    $0.78  \pm 0.11$   &     1.00 \\
$116.54  \pm 0.01$  &  (2,  0)   &   $13.3 \pm 0.5$  &    $1.00  \pm 0.08$   &     1.00 \\
$125.06  \pm 0.03$  &  (2,  0)   &   $10.8 \pm 0.8$  &    $0.84  \pm 0.15$   &     1.00 \\
$133.35  \pm 0.02$  &  (2,  0)   &    $9.3 \pm 0.6$  &    $0.85  \pm 0.09$   &     1.00 \\
 $86.01  \pm 0.01$  &  (3,  0)   &    $3.1 \pm 0.1$  &    $0.27  \pm 0.02$   &     0.68 
\enddata
\label{appendixtable}
\tablenotetext{a}{An amplitude is measured when the peak is a resolved Lorentzian, while height is measured instead when the peak is an unresolved Sinc$^2$ function. Linewidth is not defined in the latter case.}
\tablenotetext{b}{Values of 0.99 and above are ensured to be significant.}
\end{deluxetable*}

\clearpage

\bibliography{KIC-9246715-bibliography.bib}

\end{document}